\documentclass[fleqn,usenatbib]{mnras}

\usepackage[utf8]{inputenc}
\usepackage{enumitem}
\usepackage{graphicx}
\usepackage{hyperref}
\usepackage{mathtools}
\usepackage{comment}
\usepackage{amsmath}
\usepackage{ amssymb }
\usepackage{lipsum} 
\usepackage{float}
\usepackage{cleveref}
\usepackage[caption=false]{subfig}
\usepackage{placeins}
\usepackage{booktabs}
\usepackage{subcaption}
\usepackage{comment}
\usepackage{graphicx}
\usepackage{amsmath,amssymb}
\usepackage{hyperref}
\usepackage{braket}
\usepackage{float}
\usepackage[dvipsnames]{xcolor}
\usepackage{multirow}
\usepackage{mathtools}
\usepackage{makecell}

\usepackage[caption = false]{subfig}
\usepackage{txfonts}
\usepackage[normalem]{ulem}

\title[Consistent modeling of gas and matter]
{GODMAX: Modeling gas thermodynamics and matter distribution using JAX}

\author[Pandey et al.]{Shivam Pandey\thanks{\href{mailto:sp4204@columbia.edu}{sp4204@columbia.edu}},$^{1,2}$
Jaime Salcido,$^{3}$
Chun-Hao To,$^{4,5,6}$
J.~Colin Hill,$^{1,2}$
Dhayaa Anbajagane,$^{7,8}$
\newauthor
Eric J. Baxter,$^{9}$
Ian G. McCarthy$^{3}$
\\$^{1}$Department of Physics, Columbia University, 538 West 120th Street, New York, NY, USA 10027, USA
\\$^{2}$Columbia Astrophysics Laboratory, Columbia University, 550 West 120th Street, New York, NY 10027, USA
\\$^{3}$Astrophysics Research Institute, Liverpool John Moores University, 146 Brownlow Hill, Liverpool L3 5RF, UK
\\$^{4}$Center for Cosmology and AstroParticle Physics (CCAPP), Ohio State University, Columbus, OH 43210, USA
\\$^{5}$Department of Physics, Ohio State University, Columbus, OH 43210, USA
\\$^{6}$Department of Astronomy, Ohio State University, Columbus, OH 43210, USA
\\$^{7}$Department of Astronomy and Astrophysics, University of Chicago, Chicago, IL 60637,
USA
\\$^{8}$Kavli Institute for Cosmological Physics, University of Chicago, Chicago, IL 60637, USA
\\$^{9}$Institute for Astronomy, University of Hawai`i, 2680 Woodlawn Drive, Honolulu, HI 96822, USA
}

\begin{document}

\label{firstpage}

\pagerange{\pageref{firstpage}--\pageref{lastpage}}
\maketitle

\begin{abstract}
We introduce GODMAX (Gas thermODynamics and Matter distribution using jAX), a novel code designed to calculate correlations between the cosmological matter distribution and various gas thermodynamic quantities. Utilizing the extensive \texttt{ANTILLES} suite of 200 hydrodynamical simulations with a diverse range of baryonic feedback strengths, we jointly fit the 3D profiles of total matter distribution, electron density, and electron pressure across various halo masses and redshifts. By accommodating significant variations in gas profiles expected due to baryonic feedback, solving exact hydrostatic equilibrium equation and offering flexible modeling of non-thermal pressure support, GODMAX has the capability to jointly fit all these profiles within the measurement uncertainties. This advancement enables, for the first time, robust joint analyses of multiple cosmic probes, including the kinetic and thermal Sunyaev-Zel'dovich effect, weak lensing, and X-ray observations. Furthermore, the model accurately captures correlations between the total matter power suppression due to baryonic feedback and local average thermodynamic quantities, such as the baryon fraction and integrated tSZ effect, in high-mass halos, aligning with observations from hydrodynamical simulations. Looking ahead, we forecast the expected constraints on cosmological and baryonic parameters from upcoming weak lensing catalogs from the LSST and tSZ maps from the Simons Observatory. This analysis underscores the importance of cross-correlations between weak lensing and tSZ in enhancing parameter constraints by resolving major systematic uncertainties due to baryonic physics. The GODMAX code leverages the \texttt{JAX} library, resulting in a fully differentiable halo model with native GPU compilation support.
\end{abstract}  


\section{Introduction}\label{sec:intro}

The formation of the large-scale structure (LSS) of the Universe is a multi-scale phenomenon. Large-scale gravitational fields dictate the movement of matter, forming rich structures of dark matter and baryons such as halos, filaments, sheets, and galaxies. However, active galactic nuclei (AGN) at the centers of massive galaxies and supernovae release large amounts of energy in the form of jets and high-speed galactic winds which can impact the baryon and dark matter distribution on significantly larger scales ($\sim$ mega-parsec) \citep{vanDaalen:2011:MNRAS:, Schneider:2016:JCAP:, Borrow:2020:MNRAS:, Hafen:2020:MNRAS:, Gebhardt:2023:arXiv:}.
Understanding this highly non-linear feedback process is crucial for obtaining accurate cosmological constraints, pinning down galaxy formation dynamics, and quantifying the validity of hydrodynamical simulations.

As the high-speed jets push hot gas out of halos, altering its temperature and density, it in turn affects the dark matter distribution through gravity. Consequently, the feedback effect has a correlated impact on the total matter distribution as well as the density, temperature, and pressure of the surrounding hot gas. The total matter distribution is directly probed by weak lensing of light emitted from galaxies and the cosmic microwave background (CMB). However, a poor understanding of baryonic feedback is a leading source of systematic uncertainty in cosmological analyses from current-generation weak lensing surveys and this problem will become more severe with future high-precision surveys \citep{Chisari:2019:OJAp:}.

We can simulate the effects of baryons on the total matter distribution with hydrodynamical simulations. Various simulations, each with independent hydrodynamical prescriptions, feedback mechanisms, and feedback strengths, have shown that baryonic feedback leads to the suppression of the total matter power spectrum \citep{vanDaalen:2011:MNRAS:, vanDaalen:2020, Schaye:2015:MNRAS:, Springel:2018:MNRAS:, Dave:2019:MNRAS:, McCarthy:2017:MNRAS:, Villaescusa-Navarro:2021:ApJ:, Salcido:2023:MNRAS:}. The exact amplitude and scale dependence of the power suppression vary depending on the specifics of the hydrodynamical simulations. For scales probed by current-generation surveys, this effect can be as large as 10-20\%. While running large-volume hydrodynamical simulations is computationally challenging, given the uncertainty in baryonic feedback mechanisms, a large variety of simulations is necessary to understand the impact of baryons on the matter distribution.

There are alternative methodologies to mitigate the impact of baryons. Multiple works have modified the halo model prescription (see \cite{Cooray:2002:PhR:} for a review) to calculate the total matter power spectrum. These modifications include adding various components such as gas and stars \citep{Rudd:2008:ApJ:, Guillet:2010:MNRAS:, Debackere:2020:MNRAS:, Semboloni:2013:MNRAS:, Fedeli:2014:JCAP:a, Fedeli:2014:JCAP:b} or introducing parameterized freedom in the total matter distribution to account for galaxy formation effects \citep{Mead:2015:MNRAS:, Mead:2021:MNRAS:}. In \cite{Eifler:2015:MNRAS:}, principal components of the variations in matter power spectrum suppression were extracted from existing hydrodynamical simulations, which can be marginalized over in data analysis. However, to accurately pinpoint the baryonic effects, we need a probe more sensitive to baryonic processes than the total matter distribution.

The Sunyaev-Zel'dovich (SZ) effect \citep{Sunyaev:1972:CoASP:}, which includes the thermal SZ (tSZ) effect and the kinetic SZ (kSZ) effect, is caused by the inverse-Compton scattering of CMB photons on their path from the surface of last scattering to us. The tSZ effect, resulting from the random thermal motion of gas, and the kSZ effect, stemming from the coherent bulk motion of gas, are key to probing baryonic feedback as they directly probe the thermodynamics of the surrounding hot gas. Notably, as the SZ effects are a product of scattering processes, they are not subject to redshift dimming, making it easier to probe the higher-redshift Universe compared to X-ray observations of baryons. Therefore, the cross-correlations of weak lensing and SZ surveys offer a window to study the joint impact of baryonic feedback on both the matter distribution and baryon thermodynamics. Moreover, this multi-probe analysis of SZ and weak lensing is sensitive to a large dynamical range of halo masses and redshifts, enabling constraints on the variation of feedback over time and across LSS environments.

However, a joint analysis of weak lensing and SZ requires a model that can predict the joint impact of baryonic feedback on both the matter distribution and gas thermodynamics. Various studies have developed models for the thermodynamics of baryons in the intergalactic and intracluster medium \citep{Komatsu:2001:MNRAS:, Ostriker:2005:ApJ:, Nagai:2007:ApJ:, Shaw:2010:ApJ:, Osato:2023:MNRAS:}. However, these models assume that the underlying matter distribution, which sets the gravitational potential for gas thermodynamics, remains unchanged by baryonic processes. Baryonic effects can, nevertheless, alter the total gravitational potential, for instance, by ejecting gas due to feedback or by the cooling of gas to form stars and galaxies. Therefore, a joint model must account for this change and its impact on baryonic thermodynamics.

As mentioned, \cite{Schneider:2016:JCAP:} and \cite{Schneider:2019:JCAP:} developed a `baryonification' model to describe the impact of baryons on the dark matter distribution. They divided the total matter inside a halo into stellar, gas, and dark matter components. By using conservation of mass and employing physical and empirically motivated shapes of profiles for these components, they modeled their correlated distribution. However, for their joint analysis with SZ surveys, a joint prediction for the thermodynamics of gas is also required. In this study, we introduce \texttt{GODMAX}, a general framework for connected modeling of baryonic thermodynamics and the total matter distribution, building upon the baryonification model. We make minimal assumptions on the distributions of individual matter components and baryon thermodynamics, except for imposing physical conservation laws.

With the \texttt{GODMAX} model, we aim to analytically study observations made in hydrodynamical simulations. For instance, previous analyses have shown that the total matter power suppression on small scales is correlated with the mean baryon fraction \citep{vanDaalen:2020, Salcido:2023:MNRAS:} and mean integrated tSZ signal \citep{Pandey:2023:MNRAS:} of high-mass halos ($M \sim 10^{14} M_{\odot}/h$), which we replicate analytically in this study. We also examine the correlation of the integrated tSZ signal with secondary properties of halos as discussed in \cite{Lee:2022:MNRAS:, Hadzhiyska:2023:MNRAS:, Baxter:2024:MNRAS:}.

\cite{Mead:2020:A&A:} describes another joint model for the total matter distribution and gas pressure. However, this model uses simple approximations for gas density and pressure profiles, imposing that they follow a polytropic form \citep{Komatsu:2001:MNRAS:}. It is known, however, that this model is insufficient for accurately modeling the pressure and density of low-mass halos \citep{Capelo:2012:MNRAS:, Battaglia:2012:ApJ:b}. With a significantly more flexible gas distribution, solving accurate hydrostatic equilibrium equations, and more realistic modeling of non-thermal pressure support, we find that the \texttt{GODMAX} model can jointly describe not only the total matter distribution and gas pressure but also gas density across a wide range of halo masses, redshifts, and hydrodynamical simulations. This capability will enable a joint analysis of all correlations constructed out of weak lensing of galaxies and CMB, tSZ, kSZ, and X-ray surveys.

Excitingly, the sensitivity of SZ and weak lensing measurements is expected to increase dramatically in the near future, thanks to high-sensitivity CMB surveys \citep{Benson:2014:SPIE:, Henderson:2016:JLTP:, Ade:2019:JCAP:, Abazajian:2016:arXiv:,CMB-S4:2019} and galaxy surveys \citep{TheLSSTDarkEnergyScienceCollaboration:2018:arXiv:, EuclidCollaboration:2020:A&A:, Spergel:2015:arXiv:}. Therefore, the time is ripe for the development, validation, and application of a joint model of the matter distribution and gas thermodynamics.

For computational efficiency, the \texttt{GODMAX} model has been implemented using the \texttt{JAX}\footnote{\url{https://jax.readthedocs.io}} library. This offers automatic differentiation (\texttt{autodiff}) of most \texttt{NumPy} and native \texttt{Python} functions, out-of-the-box parallelization schemes (\texttt{vmap}), and just-in-time (\texttt{jit}) compilation for both CPU and GPU. The \texttt{autodiff} functionality makes it easy to interface with efficient sampling schemes like Hamiltonian Monte Carlo \citep{HMCDuane, HMCneal} and Langevin Monte Carlo \citep{MolecularDynamics, Robnik:2023:arXiv:}. To our knowledge, this is the first implementation of hydrodynamical as well as halo model calculations in \texttt{JAX}.

The outline of the paper is as follows: in Section~\ref{sec:model} we describe our joint model of the matter distribution and gas thermodynamics, as well as calculations of the observable two-point correlations between weak lensing and tSZ surveys. In Section~\ref{sec:simulated_datasets}, we detail the hydrodynamical simulations used to validate the \texttt{GODMAX} code and the specifications of future LSS and CMB surveys employed to forecast the benefits of joint modeling of weak lensing and tSZ. In Section~\ref{sec:results}, we present our results, and in Section~\ref{sec:summary}, we conclude.

\section{Modeling}\label{sec:model}
The aim of this study is to construct a general model for the distribution of relevant matter components as well as gas thermodynamics within the halo model of LSS. We adopt the notations from \cite{Schneider:2019:JCAP:} and \cite{Giri:2021:JCAP:} to describe the distributions of matter components, but with a more flexible parameterization to account for their dependence on halo mass, redshift, and concentration as observed in simulations (e.g. Eq.~\ref{eq:rho_gas} and Eq.~\ref{eq:theta_ej_co}). 
% We then augment this model to consistently predict the thermodynamics of hot gas within the halo.
Then conditioned on the dark matter and baryon distribution within a halo, we consistently predict the thermodynamics of the electrons in the halo, which are then used to predict the SZ observables. 

Our starting point is a universe where all matter (including baryons) behaves as collisionless dark matter. Governed solely by gravity to form structures, dark matter collapses into halos. The density profiles of these halos can be well approximated by a truncated Navarro-Frenk-White (NFW) profile \citep{Navarro:1996:ApJ:, Baltz:2009:JCAP:}. For any halo of mass $M_{\rm 200c}$, at redshift $z$, and with concentration $c_{\rm 200c}$, the truncated NFW profile is given by:
\begin{equation}\label{eq:rho_nfw}
    \rho_{\rm nfw}(x)=\frac{\rho_{\rm nfw,0}}{x(1+x)^2}\frac{1}{(1+y^2)^2}\,,
\end{equation}
where $x = r/r_{\rm s}$ and $y = r/r_{\rm t}$. The scale radius, $r_{\rm s}$, is defined as $r_{\rm s} = r_{\rm 200c}/c_{\rm 200c}$, and we set the truncation radius, $r_{\rm t}$, to be $r_{\rm t} = 4\times r_{\rm 200c}$ \citep{Oguri:2011:MNRAS:, Schneider:2019:JCAP:}. The normalization constant, $\rho_{\rm nfw,0}$, is fixed such that the total mass inside $r_{\rm 200c}$ integrates to $M_{\rm 200c}$. Note that the spherical overdensity radius, $r_{200c}$, of a halo at redshift $z$ is defined such that the average enclosed density within a sphere of radius $r_{\rm 200c}$ is equal to $200$ times the critical density of the Universe, $\rho_c(z)$: $M_{\rm 200c} = (4\pi/3) \, 200 \, r_{\rm 200c}^3 \, \rho_c(z)$, where the spherical overdensity mass, $M_{\rm 200c}$, is the mass within $r_{\rm 200c}$. Note that we suppress explicit redshift dependence in the equations below for the sake of clarity. Moreover, unless otherwise stated, we fix the cosmological parameters to the values mentioned in Table~\ref{tab:params_all}. Note that we assume neutrinos are massless in this work and leave the exploration with massive neutrinos to future work.

We now turn to how various baryonic components, having properties different from collisionless dark matter, populate this halo. Unlike dark matter, baryons can interact through forces other than gravity, enabling them to cool and form stars, galaxies, and AGNs. High-energy astrophysics complicates their profiles, and in the following sections, we model each baryonic and dark matter component separately.

\subsection{Matter distribution}
The total density of a halo containing both dark matter and baryons ($\rho_{\rm dmb}$) can be split into three major components: stars in the central galaxy ($\rho_{\rm cga}$), hot gas ($\rho_{\rm gas}$), and collisionless matter ($\rho_{\rm clm}$):
\begin{equation}\label{eq:rho_dmb}
    \rho_{\rm dmb}(r) = \rho_{\rm cga}(r) + \rho_{\rm gas}(r) + \rho_{\rm clm}(r),
\end{equation}
where collisionless matter includes both dark matter and stars in satellite galaxies. Given that we are describing the same halo as in Eq.~\ref{eq:rho_nfw}, their total masses should be equivalent. We define $M_{\Gamma}(<r) = 4\pi \int_0^r ds \, s^2 \, \rho_{\Gamma}(s)$ as the total mass inside a radius $r$ for any component $\Gamma \in \{\rm nfw, dmb, cga, gas, clm\}$. Then, for a sufficiently large value of $r$, we have a consistency relation indicating that the total mass of any halo $M_{\rm tot}$ is conserved:
\begin{equation}
    M_{\rm tot} = M_{\rm cga}(< \infty) + M_{\rm gas}(< \infty) + M_{\rm clm}(< \infty) = M_{\rm nfw}(< \infty).
\end{equation}
We now present the motivated functional forms of these three components comprising the dark matter and baryon profile of the halo.

\subsubsection{Central galaxy stellar profile}\label{sec:cga_profile}
The stellar density profile of the central galaxy is described using a power-law profile with an exponential cutoff \citep{Mohammed:2014:arXiv:}:
\begin{equation}
    \rho_{\rm cga}(r)=\frac{f_{\rm cga}M_{\rm tot}}{4\pi^{3/2}R_{\rm h}}\frac{1}{r^2}\exp\left[-\left(\frac{r}{2R_{\rm h}}\right)^2\right],
\end{equation}
where $R_{\rm h}=0.015\,r_{\rm 200c}$ is the stellar half-light radius, and $f_{\rm cga}$ is the total abundance of stars in the central galaxy. We parameterize $f_{\rm cga}$ as:
\begin{equation}
    f_{\rm cga}(M_{\rm 200c}) = A\left(\frac{M_1}{M_{\rm 200c}}\right)^{\mu_{\rm cga}}.
\end{equation}
Following \citet{Moster:2013:MNRAS:}, we set $A=0.09$, $M_1=2.5\times10^{11}$ $M_{\odot}/h$ and $\mu_{\rm cga} = 0.6$.  

In a subsequent subsection (Section~\ref{sec:clm_profile}), we will describe the modeling of stars in satellite galaxies with total abundance of $f_{\rm sga}$. The total stellar fraction is denoted as $f_{\rm star} = f_{\rm cga} + f_{\rm sga}$, where $f_{\rm star}$ is similarly parameterized:
\begin{equation}\label{eq:fstar}
    f_{\rm star}(M_{\rm 200c}) = A\left(\frac{M_1}{M_{\rm 200c}}\right)^{\mu_{\rm star}},
\end{equation}
and we impose $\mu_{\rm star} < \mu_{\rm cga}$ to ensure that $f_{\rm star} > f_{\rm cga}$. We note that since we only analyze the total matter distribution (in $r/r_{\rm 200c} > 0.05$) and gas thermodynamics in this work, our sensitivity to the stellar profile is quite low. Therefore, we only treat $\mu_{\rm star}$ as a free parameter where required.

\subsubsection{Gas profile}\label{sec:gas_profile}
Following \cite{Giri:2021:JCAP:}, we parameterize the gas density profile as:
\begin{equation}\label{eq:rho_gas}
\rho_\mathrm{gas}(r) = \frac{\rho_{\rm gas, 0}}{\left[1+ \left(\frac{r}{\theta_{\rm co} r_\mathrm{200c}}\right) \right]^{\beta} \left[1+ \left(\frac{r}{\theta_\mathrm{ej}r_\mathrm{200c}}\right)^\gamma \right]^{\frac{\delta-\beta}{\gamma}}},
\end{equation}
where the parameters $\theta_{\rm co}$ and $\theta_{\rm ej}$ control the core and ejection radii of the gas, respectively. The parameters $\beta, \gamma$, and $\delta$ control its slope. These parameters account for the impact of baryonic feedback which typically ejects the gas from inside the halo to its outskirts, resulting in a profile that can depart significantly relative to the NFW profile by becoming cored in the center and with shallower slope in the outskirts. The normalization factor $\rho_{\rm gas, 0}$ is fixed by requiring that $M_{\rm gas}(< \infty) = f_{\rm gas} M_{\rm tot}$, where $f_{\rm gas} = \Omega_{\rm b}/\Omega_{\rm m} - f_{\rm star}$ is the universal gas fraction. 

We find that the parameters $\theta_{\rm co}, \theta_{\rm ej}, \beta$, and $\gamma$ are the most significant in impacting both the total matter distribution and gas thermodynamics. Ideally, all of these parameters should depend on halo mass, redshift, concentration, and other secondary properties. However, due to the degeneracy in their impact on the gas profile, we vary only some of these parameters in this study. We expect the gas profile to be shallower than the NFW profile in the outskirts as the baryons are ejected due to feedback with a slope that depends on the efficiency of feedback, which evolves with halo mass \citep{Eckert:2016:A&A:}. We parameterize the evolution of $\beta$ with mass as
\begin{equation}\label{eq:beta_gas}
    \beta = \frac{3(M_{\rm 200c}/M_{\rm c})^{\mu_{\beta}}}{1+(M_{\rm 200c}/M_{\rm c})^{\mu_{\beta}}} \ ,
\end{equation}
where $M_c$ controls the mass below which the gas profile becomes shallower than the NFW profile. We further evolve the parameter $M_c$ with redshift as
\begin{equation}\label{eq:Mc_gas}
    M_{\rm c} = M_{\rm c,0} (1 + z)^{\nu_{M_{\rm c}}}
\end{equation}
Here, $M_{\rm c,0}$, $\mu_{\beta}$, and $\nu_{M_{\rm c}}$ are treated as free parameters. While the parameter $\gamma$ could also similarly evolve with redshift and halo mass, we only vary its amplitude to reduce the degeneracy in the parameter space.

The evolution of $\theta_{\rm ej}$ and $\theta_{\rm co}$ is parameterized as
\begin{equation}\label{eq:theta_ej_co}
    \theta_{\rm ej/co} = \theta_{\rm ej/co,0} \bigg(\frac{M_{\rm 200c}}{M_{\rm ej/co}}\bigg)^{\mu_{\rm ej/co}} \, (1 + z)^{\nu_{\rm ej/co}} \, \bigg(\frac{1}{c_{\rm 200c}}\bigg)^{\eta_{\rm ej/co}}.
\end{equation}
Note that here we explicitly include a dependence on concentration. To first order, the gas profile concentration is expected to be positively correlated with the dark matter concentration \citep{Komatsu:2001:MNRAS:}. We can achieve this by increasing the value of $\eta_{\rm ej/co}$.

\subsubsection{Collisionless matter profile}\label{sec:clm_profile}
We adopt the approach of \cite{Schneider:2016:JCAP:} and \cite{Schneider:2019:JCAP:} to model the relaxation of collisionless matter due to baryonic components. The gravitational effects of the gas and central galaxy stellar profile cause the shells of collisionless matter to either contract or expand, preserving angular momentum under the adiabatic relaxation condition. Assuming the initial radius of a shell is $r_i$ and its final radius is $r_f$, simulations have demonstrated that this relaxation effect is accurately captured by \citep{Abadi:2010:MNRAS:, Teyssier:2011:MNRAS:}:
\begin{equation}\label{eq:zeta}
    \zeta = \frac{r_f}{r_i} - 1 = a_{\zeta} \bigg[ \bigg( \frac{M_i}{M_f} \bigg)^{n_{\zeta}} - 1 \bigg],
\end{equation}
where we set $a_{\zeta}=0.3$ and $n_{\zeta}=2$ following \citet{Abadi:2010:MNRAS:}. The masses inside the shells are determined by:
\begin{equation}
\begin{split}
M_i&= M_{\rm nfw}(<r_i),\\
M_f&= f_{\rm clm}M_{\rm nfw}(<r_i)+ M_{\rm cga}(<r_f)+M_{\rm gas}(<r_f),
\end{split}
\end{equation}
where $f_{\rm clm} = (\Omega_{\rm m} - \Omega_{\rm b})/\Omega_{\rm m} + f_{\rm star} - f_{\rm cga}$ represents the fraction of collisionless matter. Eq.~\ref{eq:zeta} is then iteratively solved to obtain $\zeta$. Using this, we can express the relaxed collisionless matter density as:
\begin{equation}
    \rho_{\rm clm}(r) = \frac{f_{\rm clm}}{\zeta^3}\rho_{\rm nfw}(r/\zeta).
\end{equation}

\subsection{Gas thermodynamics}\label{sec:gas_thermodynamics}

Assuming hydrostatic equilibrium, the total pressure is given by:
\begin{equation}
    \rho_{\rm gas}^{-1} \, \frac{dP_{\rm tot}}{dr} = -G \frac{M_{\rm dmb}(<r)}{r^2},
\end{equation}
where $M_{\rm dmb}(<r) = 4\pi \int_0^r s^2 \rho_{\rm dmb}(s) ds$. This condition is expected in a cluster with no ongoing active merger. Given a set of parameters, we can obtain $M_{\rm dmb}(<r)$ and $\rho_{\rm gas}$, and then solve the above equation to obtain the total pressure $P_{\rm tot}$. This total pressure is composed of thermal ($P_{\rm th}$) and non-thermal ($P_{\rm nt}$) pressure components. 

The fraction of non-thermal pressure support is expressed as \citep{Shaw:2010:ApJ:, Osato:2023:MNRAS:}:
\begin{equation}\label{eq:Pnt}
  R_\mathrm{nt} = \frac{P_\mathrm{nt}}{P_\mathrm{tot}} = \alpha_\mathrm{nt} f(z)
  \left( \frac{r}{r_{\rm 200c}} \right)^{n_\mathrm{nt}},
\end{equation}
where $\alpha_{\rm nt}$ controls the amplitude of non-thermal pressure support, and $f(z)$ governs its redshift evolution. We parameterize this function as:
\begin{equation}
  f (z) = \mathrm{min} \left[ (1+z)^{\beta_\mathrm{nt}},
  (f_\mathrm{max}-1) \tanh (\beta_\mathrm{nt} z) + 1 \right] ,
\end{equation}
where $f_\mathrm{max} = 6^{-n_\mathrm{nt}} / \alpha_\mathrm{nt}$
ensures that $R_\mathrm{nt} < 1$ for pressures in the relevant radial range of $0 < r < 6r_{\rm 200c}$ \citep{Shaw:2010:ApJ:}.
Thus, the thermal pressure $P_\mathrm{th}$ is given by:
\begin{equation}
  P_\mathrm{th} = P_\mathrm{tot} \times \mathrm{max}
  \left[ 0, 1 - R_\mathrm{nt} \right] .
\end{equation}

The pressure of free electrons from the thermal pressure of gas is calculated as:
\begin{equation}\label{eq:Pe}
    P_e(r) = \frac{2(X_{\rm H} + 1)}{(5X_{\rm H} + 3)} P_{\rm th},
\end{equation}
where $X_{\rm H} = 0.76$ is the primordial hydrogen mass fraction.

Moreover, with the gas density profile, the free electron density profile can be determined as:
\begin{equation}\label{eq:ne}
    n_e(r) = \frac{\rho_{\rm gas}(r)}{m_p \mu_e},
\end{equation}
where $m_p$ is the proton mass and $\mu_e$ is the mean molecular weight per electron (assuming metal abundances of 0.3 solar), fixed at $\mu_e = 1.17$ \citep{Anders:1989:GeCoA:}. 

This enables us to solve for the electron temperature profile using the relation $P_e = k_{\rm B} n_e T_e$, where $k_{\rm B}$ is the Boltzmann constant.

Note that the kSZ effect is directly sensitive to the \textit{physical} electron density profile and the tSZ signal is sourced by the \textit{physical} electron pressure profile. We perform this conversion from comoving to the physical coordinates when predicting relevant SZ observables.

When comparing with observations, it is often more convenient to compare with the mean pressure and gas density inside the halo boundary. We calculate the average baryon fraction $f_{\rm b}$ inside a halo by integrating the gas and stellar distributions out to $r_{\rm 200c}$. Similarly, the integrated tSZ signal within $r_{\rm 200c}$ can be obtained by integrating the gas pressure as:
\begin{equation}\label{eq:Y3D}
    Y^{\rm 3D}_{\rm 200c} = 4\pi \int_0^{r_{\rm 200c}} ds \, s^2 P_e(s).
\end{equation}
Note that this integrated profile over 3D volume is not directly observable from the tSZ measurements, which probe the integrated pressure along the line of sight. However, this can be computed from the inferred 3D pressure profiles from the analysis of cross-correlation data between halos and tSZ, and assuming spherical symmetry of the stacked measurements (e.g.,~\citet{Hill:2018:PRD,Pandey:2022}).

For a cluster in virial equilibrium, ignoring the effects of baryonic feedback and non-thermal pressure, the self-similar expectation of the 3D integrated tSZ signal can be derived as \citep{Battaglia:2012:ApJ:a}:
\begin{equation}
\label{eq:y_ss}
    Y^{\rm 3D}_{\rm SS} = 97.6 \, h^{-1}_{70} \,\bigg( \frac{M_{\rm 200c}[ M_{\odot}]}{10^{15} h^{-1}_{70}[M_{\odot}]} \bigg)^{5/3} \frac{\Omega_{\rm b}}{0.043} \frac{0.25}{\Omega_{\rm m}} \, {\rm kpc^2},
\end{equation}
where $h_{70} = h/0.7$.

\subsection{Matter power spectrum}\label{sec:Pmm}
Using the halo model framework (see \citealt{Cooray:2002:PhR:} for a review), we can express the matter power spectrum as a sum of the 1-halo and 2-halo terms. The 1-halo term can be written as:
\begin{equation}
    P^{\rm mm; 1h}_{\rm dmb/nfw}(k,z) = \int_{M_{\rm min}}^{M_{\rm max}} dM \frac{dn}{dM} \int_{c_{\rm min}}^{c_{\rm max}} dc \, p(c|M) \, u^2_{\rm dmb/nfw}(k, z, M, c),
\end{equation}
where $u_{\rm dmb/nfw}$ is the Fourier-space profile of the total matter distribution and $dn/dM$ is the halo mass function, for which we use the \cite{Tinker:2010:ApJ:} fitting function. We choose the integral limits as $M_{\rm min} = 10^{11.0}$ $M_\odot/h$, $M_{\rm max} = 10^{16.0}$ $M_\odot/h$, $c_{\rm min} = 2.0$, and $c_{\rm max} = 8.0$, ensuring convergence of the integral. The term $p(c|M)$ encapsulates the distribution of halo concentrations at a given halo mass, well approximated by a log-normal distribution \citep{Bullock:2001:ApJ:}:
\begin{equation}
    p(c|M) dc = \frac{d\ln c}{\sqrt{2\pi c}}\exp \Bigg[-\frac{\ln^2 (c/\bar{c}(M))}{2\sigma^2_{\ln c}} \Bigg],
\end{equation}
where $\bar{c}(M)$ is the mean concentration-mass relationship, for which we use the fitting function described in \cite{Diemer:2015:ApJ:}.

The Fourier transform of the profile of the total matter distribution, either in the gravity-only case (NFW) or the baryonic case (DMB), is given by:
\begin{equation}\label{eq:uk_dmb}
    u_{\rm dmb/nfw}(k,z,M,c) = \int_0^{\infty} dr \, 4\pi r^2 \frac{\sin(kr)}{kr} \frac{\rho_{\rm dmb/nfw}(r)}{\bar{\rho}_{\rm m, 0}},
\end{equation}
where $\bar{\rho}_{\rm m, 0}$ is the mean comoving matter density of the Universe.

The 2-halo term is given by:
\begin{equation}
    P^{\rm mm; 2h}_{\rm dmb/nfw}(k,z) = (b^{\rm m}_{\rm dmb/nfw}(k, z))^2 \, P_{\rm lin}(k,z),
\end{equation}
where $b^{\rm m}_{\rm dmb/nfw}(k, z)$ is the scale-dependent bias of the total matter field, and $P_{\rm lin}(k,z)$ is the linear matter power spectrum for any given cosmology. We use \cite{Campagne:2023:OJAp:} to calculate $P_{\rm lin}$ within the \texttt{JAX} framework using the Eisenstein-Hu approximation \citep{Eisenstein:1998:ApJ:}. Since we limit ourselves to modeling correlations between projected fields such as tSZ or weak lensing, this is a sufficient approximation. We note that more accurate implementations of the linear matter power spectrum in \texttt{JAX} have been developed \citep{Hahn:2023:arXiv:}, but interfacing with these implementations is left for a future study.

Note that the consistency relation of mass conservation requires that the average bias of total matter on large scales should approach unity. However, achieving this requires setting $M_{\rm min} \rightarrow 0$, which makes the integral slow to converge. Therefore, we follow the approach outlined in \cite{Cacciato:2012:MNRAS:, Schmidt:2016:PhRvD:, Mead:2020:A&A:, Bolliet:2023:JCAP:} to estimate the average bias in the case of a finite $M_{\rm min}$. In this approach, the effective large-scale bias can be expressed as:
\begin{equation}\label{eq:bm_dmb} 
    b^{\rm m}_{\rm dmb/nfw} = I_{\rm dmb/nfw}(M_{\rm min}) + A(M_{\rm min}),
\end{equation}
where $I_{\rm dmb/nfw}$ is the bias calculated from all halos with masses above $M_{\rm min}$, and $A(M_{\rm min})$ adds the contribution from lower mass halos such that on very large scales, mass is conserved and the effective bias is 1. The term $I_{\rm dmb/nfw}$ can be estimated as:
\begin{equation}     
    I_{\rm dmb/nfw} = \int_{M_{\rm min}}^{M_{\rm max}} dM \frac{dn}{dM} \int_{c_{\rm min}}^{c_{\rm max}} dc \, p(c|M) \, b^{\rm lin}_{\rm halo}(M) \, u_{\rm dmb/nfw},
\end{equation}
where $b^{\rm lin}_{\rm halo}(M)$ is the linear halo bias, for which we use the fitting function from \cite{Tinker:2010:ApJ:}. The term $A(M_{\rm min})$ can then be estimated from the large-scale limit ($k\to 0$) of $I_{\rm dmb/nfw}$ as:
\begin{equation}
    A(M_{\rm min}) = 1 - I_{\rm dmb/nfw}(M_{\rm min}, k\to 0).
\end{equation}

Therefore, the final matter power spectrum, whether in the gravity-only case or including baryonic effects, can be written as:
\begin{equation}
    P^{\rm mm}_{\rm dmb/nfw}(k,z) = P^{\rm mm; 1h}_{\rm dmb/nfw}(k,z) + P^{\rm mm; 2h}_{\rm dmb/nfw}(k,z).
\end{equation}
The impact of baryonic processes on the matter distribution is typically captured using the matter power suppression, which is simply the ratio $P^{\rm mm}_{\rm dmb}/P^{\rm mm}_{\rm nfw}$.

\subsection{Correlation between weak lensing and SZ}\label{sec:datavector_modeling}

Note that from observations, we only have access to projected matter fields through weak lensing observations. Similarly, the observations of SZ effects are also projected fields since they encode the impact of baryons on the CMB photons along the line-of-sight back to the surface of last scattering. We can express the correlation between any two cosmic probes $A$ and $B$, with their redshift distributions encoded by indices $i$ and $j$, as a sum of 1-halo and 2-halo terms in projected multipole space. The 1-halo term is given by:
\begin{multline}\label{eq:Cl1h}
C^{i j}_{A B; \textrm{1h}}(\ell) = \int_{z_{\rm{min}}}^{z_{\rm{max}}} dz \frac{dV}{dz d\Omega} \int_{M_{\rm{min}}}^{M_{\rm{max}}} dM \frac{dn}{dM} \int_{c_{\rm min}}^{c_{\rm max}} dc \, p(c|M) \, \\  \bar{u}^{i}_{A}(\ell,M,z) \times \ \bar{u}^{j}_{B}(\ell,M,z),
\end{multline}
where $dV$ is the cosmological volume element, $d\Omega$ is the solid angle formed by that element, and $\bar{u}^{i}_{A}(\ell,M,z)$ and $\bar{u}^{j}_{B}(\ell,M,z)$ are the multipole-space kernels of any general probes $A$ and $B$, which we will describe below.

The 2-halo term is given by:
\begin{equation}\label{eq:Cl2h}
C^{ij}_{AB;\textrm{2h}}(\ell) = \int_{z_{\rm{min}}}^{z_{\rm{max}}} dz \frac{dV}{dz d\Omega} b_{A}^{i}(\ell,z) \ b^{j}_{B}(\ell,z) \ P_{\rm{lin}}(k,z),
\end{equation}
where ${b}^{i}_{A}(\ell,z)$ and ${b}^{j}_{B}(\ell,z)$ are the effective large-scale biases of the cosmic probes $A$ and $B$.

We focus here on correlations constructed between tSZ (labeled with the Compton-$y$ parameter) and weak lensing (labeled with convergence $\kappa$).

\subsubsection{tSZ observable}
The multipole-space kernel of the Compton-$y$ parameter is related to the pressure profile of hot electrons ($P_e$) as follows~\citep{Komatsu:2002:MNRAS:,HillPajer:2013:PRD}:
\begin{multline}\label{eq:uyl}
\bar{u}^{j}_{y}(\ell,z,M,c) = b^{j}(\ell) \, \frac{4\pi r_{200c}}{\ell^2_{200c}} \frac{\sigma_T}{m_e c^2} \int_{x_{\rm{min}}}^{x_{\rm{max}}} dx \ x^2 \ P_e(x,z,M,c) \\ \times \frac{\sin(\ell x/l_{200c})}{\ell x/l_{200c}},
\end{multline}
where $x = r/r_{\rm 200c}$, $r$ is the radial distance, $l_{200c} = D_A/r_{\rm 200c}$ with $D_A$ being the angular diameter distance to redshift $z$, and $P_e$ is the electron pressure as derived in Eq.~\ref{eq:Pe}.
The term $b^{j}(\ell) = \exp{[-\ell(\ell + 1) \sigma_j^2/2]}$ captures the beam of experiment $j$, here assumed to be Gaussian for simplicity. Note that $\sigma_j = \theta^{{\rm FWHM},j}/\sqrt{8\ln 2}$, and we fix $\theta^{\rm FWHM} = 1.0$ arcmin (see Section~\ref{sec:forecast_specification}).

The effective tSZ bias $b^j_{y}$ can be expressed as:
\begin{equation}\label{eq:byl}
b^j_{y}(\ell,z) = \int_{M_{\rm{min}}}^{M_{\rm{max}}} dM \ \frac{dn}{dM} \int_{c_{\rm min}}^{c_{\rm max}} dc , p(c|M) , \bar{u}^j_{y}(\ell) b^{\rm lin}_{\rm halo},
\end{equation}
where $b^{\rm lin}_{\rm halo}$ is the linear halo bias.

\subsubsection{Weak lensing observable}
The effective multipole-space kernel of convergence can be related to the total-matter kernel as: 
\begin{equation}\label{eq:ukl}
    \bar{u}_{\kappa}^{i}(\ell,z,M,c) = \frac{W^{i}_{\kappa}(z)}{\chi^2}  u_{\rm{dmb}}(k,z,M,c),
\end{equation}
where $k = (\ell + 1/2) / \chi$, $\chi$ is the comoving distance to redshift $z$, $u_{\rm{dmb}}$ is the profile of the matter distribution in Fourier space (as defined in Eq.~\ref{eq:uk_dmb}), and $W^{i}_{\kappa}(z)$ is the lensing efficiency. The lensing efficiency is given by:
\begin{equation}
    W^{i}_{\kappa}(z) = \frac{3 H_0^2 \Omega_{\rm m}}{2 c^2} \frac{\chi}{a(\chi)} \int_{\chi}^{\infty} d\chi' n^{i}_{\kappa}(z(\chi')) \frac{dz}{d\chi'}\frac{\chi' - \chi}{\chi'}.
\end{equation}
Here, $n^{i}_{\kappa}$ represents the normalized redshift distribution of the source galaxies corresponding to the tomographic bin $i$.

For the two-halo term, the effective large-scale bias can be expressed as 
\begin{equation}\label{eq:bkl}
b_{\kappa}^{i}(\ell,z) = \frac{W^{i}_{\kappa}(z)}{\chi^2} \times b^{\rm m}_{\rm dmb}(k,z),
\end{equation}
where $k = (\ell + 1/2) / \chi$ and $b^{\rm m}_{\rm dmb}$ is the 3D scale-dependent bias of the matter field, as described in Eq.~\ref{eq:bm_dmb}.

\begin{table}
\centering 
% \resizebox{\textwidth}{!}
\tabcolsep=0.16cm
\begin{tabular}{|c| c c c|}
\hline
% \hline
Type & Parameter & Fiducial, Prior & Equation  \\ \hline
% & & & \\

% 
\multirow{16}{*}{\shortstack[c]{Baryonic\\ Parameters}} & \multicolumn{3}{c|}{\textbf{Gas Profile}} \\   & $\theta_{\rm ej, 0}$ & 4.0,$\mathcal{U}[1.0,8.0]$ & Eq.~\ref{eq:theta_ej_co} \\
& $\mu_{\rm ej}$ & 0.0,$\mathcal{U}[-5.0, 0.0]$ & Eq.~\ref{eq:theta_ej_co}  \\
& $\nu_{\rm ej}$ & 0.0,$\mathcal{U}[-5.0,5.0]$ & Eq.~\ref{eq:theta_ej_co} \\  
& $\theta_{\rm co, 0}$ & 0.1,$\mathcal{U}[0.0, 0.8]$   & Eq.~\ref{eq:theta_ej_co} \\
& $\gamma$ & 2.0,$\mathcal{U}[0.2,6.0]$ & Eq.~\ref{eq:rho_gas} \\  
& $\mu_{\beta}$ & 0.21,$\mathcal{U}[0.0,5.0]$ & Eq.~\ref{eq:beta_gas}  \\
& $\log_{10}(M_{\rm c,0})$ & 15.0,$\mathcal{U}[10.0,16.0]$   & Eq.~\ref{eq:Mc_gas} \\
& $\nu_{\rm M_{\rm c}}$ & -2.5,$\mathcal{U}[-6.0,6.0]$ & Eq.~\ref{eq:Mc_gas} \\  
& & & \\
\cline{2-4}
& \multicolumn{3}{c|}{\textbf{Stellar Profile}} \\  
& $\mu_{\rm star}$ & 0.3,$\mathcal{U}[0.01,0.6]$ & Eq.~\ref{eq:fstar}  \\
% & & & \\
& & & \\
\cline{2-4}
% & & & \\
& \multicolumn{3}{c|}{\textbf{Non-thermal pressure}} \\  
& $\alpha_{\rm nt}$ & 0.18,$\mathcal{U}[0.01,0.5]$ & Eq.~\ref{eq:Pnt}  \\
& & & \\
\hline
\hline
% & & & \\
% & & & \\
&  
% & & & \\
\multicolumn{3}{c|}{\textbf{Cosmology}} \\ 
\multirow{16}{*}{\shortstack[c]{Other\\ Parameters \\ (\small{Fisher} \\ \small{forecast; Section~\ref{sec:forecast}})}}
& \shortstack[c]{$\Omega_{\rm m}$}   & 0.31,$\mathcal{U}[-\infty,\infty]$ &  \\ 
& \shortstack[c]{$\Omega_{\rm b}$}   & 0.049,$\mathcal{U}[-\infty,\infty]$ &  \\ 
& \shortstack[c]{$h$}   & 0.672,$\mathcal{U}[-\infty,\infty]$ &  \\ 
& \shortstack[c]{$n_{\rm s}$}   & 0.95,$\mathcal{U}[-\infty,\infty]$ &  \\ 
& \shortstack[c]{$\sigma_8$}   & 0.81,$\mathcal{U}[-\infty,\infty]$ &  \\ 
& & & \\
\cline{2-4}
& \multicolumn{3}{c|}{\textbf{Intrinsic Alignment}} \\
& $A_{\rm IA}$ & 0.1,$\mathcal{U}[-\infty,\infty]$   &  Eq.~\ref{eq:AzIA}    \\
& $\eta_{\rm IA}$ & 0.0,$\mathcal{U}[-\infty,\infty]$   &  Eq.~\ref{eq:AzIA}   \\ 
& & & \\
\cline{2-4}
% & & & \\
& \multicolumn{3}{c|}{\textbf{Shear Calibration}} \\  
& Y1 $m^{i}$ & 0.0,$\mathcal{G}[0.0, 0.013]$ & Eq.~\ref{eq:mi1},~\ref{eq:mi2}  \\ 
& Y6 $m^{i}$ & 0.0,$\mathcal{G}[0.0, 0.003]$ & Eq.~\ref{eq:mi1},~\ref{eq:mi2}  \\ 
& & & \\

\cline{2-4}
% & & & \\
& \multicolumn{3}{c|}{\textbf{Source photo-$z$ bias}} \\  
& Y1 $\Delta_z^{i}$ & 0.0,$\mathcal{G}[0.0, 0.002]$ & Eq.~\ref{eq:Delzi}  \\ 
& Y6 $\Delta_z^{i}$ & 0.0,$\mathcal{G}[0.0, 0.001]$ & Eq.~\ref{eq:Delzi}  \\ 
% & $\Delta z_{\kappa}^{3}$ & 0.0,$\mathcal{G}[0.0, 0.011]$ & Eq.~\ref{eq:Delzi}  \\ 
% & $\Delta z_{\kappa}^{4}$ & 0.0,$\mathcal{G}[0.0, 0.017]$ & Eq.~\ref{eq:Delzi} \\ 
& & & \\
\hline 
\end{tabular}
\caption{The parameters varied, their fiducial values and their prior ranges used ($\mathcal{U}[X, Y] \equiv$ Uniform prior between $X$ and $Y$; $\mathcal{G}[\mu, \sigma] \equiv$ Gaussian prior with mean $\mu$ and standard-deviation $\sigma$) in this analysis and the equations in the text where the parameter is primarily used. Note that for the baryonic parameters, the prior range mentioned is only used when fitting the hydrodynamical simulations (Section~\ref{sec:fitting_antilles}). When performing Fisher forecasts in Section~\ref{sec:forecast}, only the shear calibration and source photo-$z$ bias parameters have an informative prior.}
\label{tab:params_all}
\end{table}

\subsubsection{Final correlation functions}
With the 1-halo and 2-halo terms calculated for probes ${A, B} \in \{y, \kappa\}$, we can express the total multipole power spectra as:
\begin{equation}\label{eq:Cl_tot}
C^{ij}_{AB}(\ell) = C^{i j}_{A B; \textrm{1h}}(\ell) + C^{ij}_{AB;\textrm{2h}}(\ell).
\end{equation}

Now, converting these correlations to angular coordinates and using the flat-sky approximation, the tSZ-weak lensing correlation can be represented using the Hankel transform as
\begin{equation}\label{eq:xigty}
\xi^{i}_{\gamma_t y}(\theta) = \int \frac{d\ell \ \ell}{2\pi} J_{2}(\ell \theta) C^{i}_{\kappa y}(\ell),
\end{equation}
where $J_{2}$ is the second-order Bessel function. Here, $i$ labels the tomographic distribution of source galaxies.

The auto-correlation of shear between any two tomographic bins $i$ and $j$ includes two components, $\xi^{ij}_{+}$ and $\xi^{ij}_{-}$, which can be calculated as:
\begin{equation}\label{eq:xiplusminus}
\xi^{ij}_{+/-}(\theta) = \int \frac{d\ell \ \ell}{2\pi} J_{0/4}(\ell \theta) C^{ij}_{\kappa \kappa}(\ell),
\end{equation}
where $J_{0}$ and $J_{4}$ are the zeroth and fourth-order Bessel functions, respectively. Note that while the transform to angular space on a curved sky can be more accurately calculated as detailed in \cite{Krause:2021:arXiv:}, we limit our forecast to a maximum angular scale of 250 arcminutes, and thus defer a more precise calculation to future studies.

\subsubsection{Instrinsic alignment}
We assume a simple non-linear linear alignment model (NLA) to describe the intrinsic alignment (IA) of source galaxies, as suggested by \cite{Bridle:2007:NJPh:}. The impact of NLA can be captured by modifying the lensing efficiency as per \cite{Krause:2017:arXiv:}:
\begin{equation}
\label{eq:CIA}
W^{i}_{\kappa}(z) \longrightarrow W^{i}_{\kappa}(z) - A\left(z\right) n^i_{\kappa}(z) \frac{dz}{d \chi}\,,    
\end{equation}
where the IA amplitude is modeled using a power-law scaling:
\begin{equation}\label{eq:AzIA}
    A(z) = -A_{\rm IA} \bigg(\frac{1+z}{1+z_0}\bigg)^{\eta_{\rm IA}} \frac{C_1 \bar{\rho}_{\rm m,0}}{D(z)},
\end{equation}
and we set $z_0 = 0.62$ and $C_1 = 5\times 10^{-14} M_{\odot}^{-1}h^{-2}{\rm Mpc}^3$ following \cite{Brown:2002:MNRAS:}, with $D(z)$ representing the linear growth factor. The parameters $A_{\rm IA}$ and $\eta_{\rm IA}$ are treated as free parameters.

\subsubsection{Observational systematics}
We model the photometric uncertainty in our source redshift distribution $n^i_{\kappa}(z)$ using shift parameters ($\Delta^i_{z}$), which modify the source redshift distributions for any tomographic bin $i$, as suggested by \cite{Krause:2017:arXiv:}:
\begin{equation}\label{eq:Delzi}
    n^i_{\kappa}(z) \rightarrow n^i_{\kappa}(z - \Delta^i_{z})
\end{equation}

The multiplicative shear bias modifies the correlations for tomographic bins $i$ and $j$ as follows:
\begin{equation}\label{eq:mi1}
    \xi^{i}_{\gamma_t y}(\theta) \rightarrow (1 + m^i) \, \xi^{i}_{\gamma_t y}(\theta)
\end{equation}
\begin{equation}\label{eq:mi2}
    \xi^{ij}_{+/-}(\theta) \rightarrow (1 + m^i) \, (1 + m^j) \, \xi^{ij}_{+/-}(\theta),
\end{equation}
where $m^i$ represents the multiplicative shear bias parameters. Both $\Delta^i_{z}$ and $m^i$ for all tomographic bins are treated as free parameters but with a Gaussian prior, as described in Table~\ref{tab:params_all}.

\subsubsection{Covariance}\label{sec:covariance}
We model the covariance, $\varmathbb{C}$, as a sum of Gaussian ($\varmathbb{C}^{\rm G}$) and connected non-Gaussian ($\varmathbb{C}^{\rm cNG}$) terms. The multi-probe covariance methodology, including the tSZ observable, is detailed in \cite{Fang:2024:MNRAS:}. We first estimate the covariance in multipole space, employing a methodology similar to that of \cite{Fang:2024:MNRAS:}, but with a few simplifications. We only model the 1-halo part of the connected 4-point function for all probes \citep{Friedrich:2021:MNRAS:, Krause:2017:MNRAS:} and ignore the contribution of super-sample covariance \citep{Osato:2021:PhRvD:}. Finally, we convert the covariance from multipole space to angular space as detailed in \cite{Krause:2017:arXiv:, Pandey:2022}. We leave improvements in our covariance estimate for future studies. Note that the covariance calculation between tSZ and weak lensing requires the expected shape noise from the weak lensing survey and component-separated noise in the auto-spectrum of tSZ maps, which are detailed in Section~\ref{sec:forecast_specification}. 

\subsubsection{Fisher forecast}\label{sec:fisher_forecast}
We estimate the constraining power of future surveys analyzing the correlations $\xi^{i}_{\gamma_t y}$ and $\xi^{ij}_{+/-}$ using the Fisher matrix formalism \citep{Fisher:1935:JSTOR:, Tegmark:1997:ApJ:}. The Fisher matrix is given by:
\begin{equation}
    \varmathbb{F}_{ab} = \sum_{A,B} \frac{\partial \xi_{AB} }{\partial p_{a}} \varmathbb{C}^{-1} \frac{\partial \xi_{AB} }{\partial p_{b}},
\end{equation}
where $\xi_{AB} \in \{\xi^{i}_{\gamma_t y}, \xi^{ij}_{+}, \xi^{ij}_{-}\}$. The partial derivatives are evaluated at the fiducial parameter values given in Table~\ref{tab:params_all}. Using the \texttt{JAX} framework, all likelihood and correlation function evaluations have \texttt{autodiff} functionality, enabling us to obtain their exact gradients with respect to all input parameters. Finally, the error on parameter $p_a$, after marginalizing over other parameters, is given by $\sigma^2(p_a) = (\varmathbb{F}^{-1})_{aa}$.

\begin{figure}
    \centering
    \includegraphics[width=0.45\textwidth]{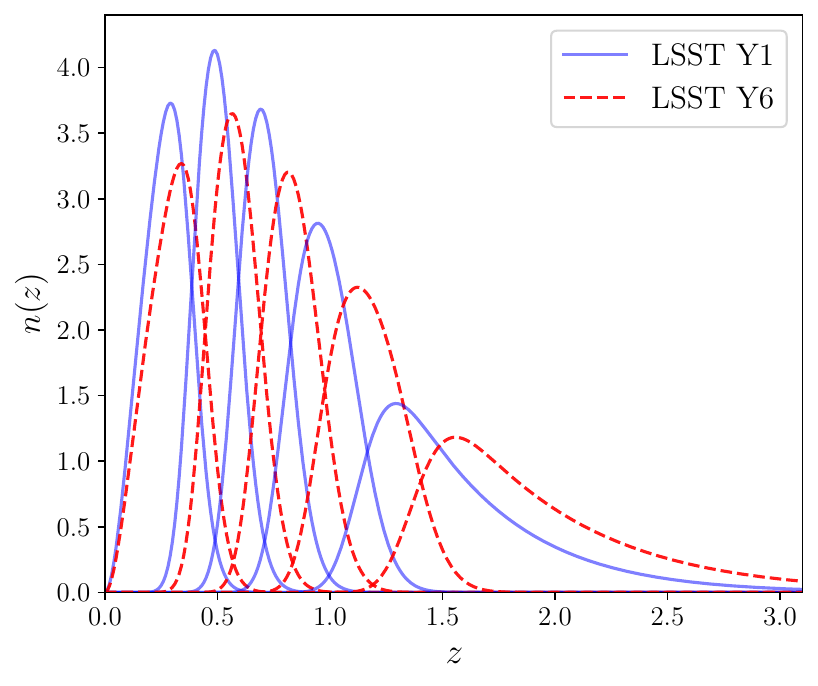}
    \caption{Simulated redshift distribution of LSST Y1 source galaxies. We divide them into five tomographic bins with same number densities.}
    \label{fig:nz_sources}
\end{figure}

\section{Simulated datasets}\label{sec:simulated_datasets}

\subsection{Hydrodynamcial simulations}
To validate the ability of our method to reproduce profiles observed in hydrodynamical simulations with varying feedback strengths, we utilize the \texttt{ANTILLES} simulation suite, as detailed in \cite{Salcido:2023:MNRAS:}. Given our goal to jointly model SZ and weak lensing correlations, it is imperative to use a simulation suite that offers a wide variation in baryonic feedback as well as diverse dynamical mass, redshift, and concentration of halos.

The \texttt{ANTILLES} suite meets these requirements, possessing a large enough volume to contain a representative sample of halos that significantly contribute to the matter power spectrum. It also provides high enough resolution to resolve the scales over which cosmological measurements are made. The suite includes 400 simulations, each with a box size of 100 Mpc/$h$ and containing $256^3$ baryon and dark matter particles. These simulations adopt a flat $\Lambda$CDM cosmology consistent with the WMAP 9-year results \citep{Hinshaw:2013:ApJS:}, and use a modified version of the \texttt{GADGET-3} smoothed particle hydrodynamics (SPH) code \citep{Springel:2005:MNRAS:} that includes a full treatment of gravity and hydrodynamics. Specifically, \texttt{ANTILLES} uses a version of \texttt{GADGET-3} that was modified for the \texttt{EAGLE} project \citep{Schaye:2015:MNRAS:}.

The simulations are designed to explore a broad range of feedback scenarios that conservatively bracket both the observed gas fractions of groups and clusters, and the observed stellar mass function. To achieve this, \texttt{ANTILLES} systematically vary the main subgrid parameters governing the efficiencies of stellar and AGN feedback. Key parameters include:
\begin{itemize}
    \item $v_w$ and $\eta_w$, which control wind velocity and mass-loading, respectively, majorly regulating galaxy and star formation efficiency \citep{DallaVecchia:2008:MNRAS:}.
    \item $n_{\rm heat}$ and $\Delta T_{\rm heat}$, which control the number and temperature increase of neighboring gas particles heated due to AGN feedback \citep{Booth:2009:MNRAS:}, with results being particularly sensitive to $\Delta T_{\rm heat}$.
    \item $n^{*}_{\rm H}$, which controls the density threshold above which the black hole accretion rate is boosted \citep{Booth:2009:MNRAS:}.
\end{itemize}

We employ a subset of 200 \texttt{ANTILLES} simulations that use the state-of-the-art \texttt{ANARCHY} SPH formulation, which includes various improvements over the standard \texttt{GADGET} SPH code \citep{Springel:2005:MNRAS:}. Comprehensive coverage of the baryonic feedback parameter space is ensured by generating 200 points using the Latin hypercube sampling technique of \cite{Deutsch:2012:JSPI:} over the five corresponding parameters. Each simulation  uses the same initial conditions, and there also exists a dark matter only simulation with the same cosmology and initial conditions. For further details on the simulations used, we refer the reader to \cite{Salcido:2023:MNRAS:}.

\subsection{Forecast Survey Specifications}
\label{sec:forecast_specification}
We apply the model presented here to imminent weak lensing and tSZ correlations expected from the Vera Rubin Observatory's LSST and the Simons Observatory (SO) within this decade. We create a mock datavector of $\xi_{\gamma_t} y$, $\xi_{+}$, and $\xi_{-}$ for five tomographic bins of source galaxies as described below and in 20 logarithmically spaced bins between 2.5 arcmin and 250 arcmin. The covariance matrix of this mock datavector is calculated analytically using the LSST and SO survey specifications as described below.

\subsubsection{Vera Rubin Observatory}
Following \cite{DESC:SRD:2018:}, we assume that LSST Year 1 (Y1) will cover a sky area of 12300 ${\rm deg}^2$. Similarly, for SO, we assume a sky area coverage of 16000 ${\rm deg}^2$. For LSST Year 6 (Y6), we anticipate the survey to cover the SO sky footprint, hence we assume $f_{\rm sky} = 0.3$ for Y1 auto- and cross-correlation analyses and $f_{\rm sky} = 0.4$ for Y6 analyses. The source samples are expected to follow a distribution given by
\begin{equation}
    n^{\rm tot}_{\kappa}(z) \propto z^2 \exp[-(z/z_{\rm pz; 0})^{\alpha_{\rm pz}}],
\end{equation}
which is normalized by the effective number density $\bar{n}^{\rm tot}_{\kappa}$. For LSST Y1, we assume $\bar{n}^{\rm tot}_{\kappa} = 11.2$, $\alpha_{\rm pz} = 0.87$, and $z_{\rm pz; 0} = 0.191$; for LSST Y6, we assume $\bar{n}^{\rm tot}_{\kappa} = 23.2$, $\alpha_{\rm pz} = 0.798$, and $z_{\rm pz; 0} = 0.178$ \citep{Fang:2022:MNRAS:, Fang:2024:MNRAS:, DESC:SRD:2018:}. This total distribution is divided into five tomographic bins with equal number densities, $\bar{n}^{i}_{\kappa} = \bar{n}^{\rm tot}_{\kappa}/5$. The normalized redshift distribution of source galaxies is shown in Fig.~\ref{fig:nz_sources}. For shape noise, we expect $\sigma_e = 0.26/{\rm component}$.

\subsubsection{Simons Observatory}
The science requirements of primary and secondary data products from SO are detailed in \cite{Ade:2019:JCAP:}. The SO large aperture telescope (LAT) will operate in frequency channels centered at 27, 39, 93, 145, 225, and 280 GHz, which will be used to construct the tSZ map. We assume a Gaussian beam with $\theta_{\rm FWHM}= 1 \, {\rm arcmin}$.\footnote{Our estimate of beam size is slightly optimistic compared to the expected 1.4 arcmin resolution of the Simons Observatory tSZ maps \citep{Ade:2019:JCAP:}. However, these specifications are used only for our forecast results (see Section~\ref{sec:forecast}), where our minimum angular scale for correlations is 2.5 arcmin. Consequently, we do not anticipate this difference in beam size to qualitatively alter our conclusions.} Detailed modeling of the instrumental noise and non-white atmospheric noise for the LAT is provided in \cite{Ade:2019:JCAP:}. We adopt the \textit{baseline} noise curve expected from the standard internal linear combination approach for separating the tSZ signal from other components in the frequency maps.\footnote{\url{https://github.com/simonsobs/so_noise_models/blob/master/LAT_comp_sep_noise/v3.1.0/SO_LAT_Nell_T_atmv1_baseline_fsky0p4_ILC_tSZ.txt}}

% \subsubsection{Mock datavector}

\begin{figure*}
    \centering
    \includegraphics[width=0.93\linewidth]{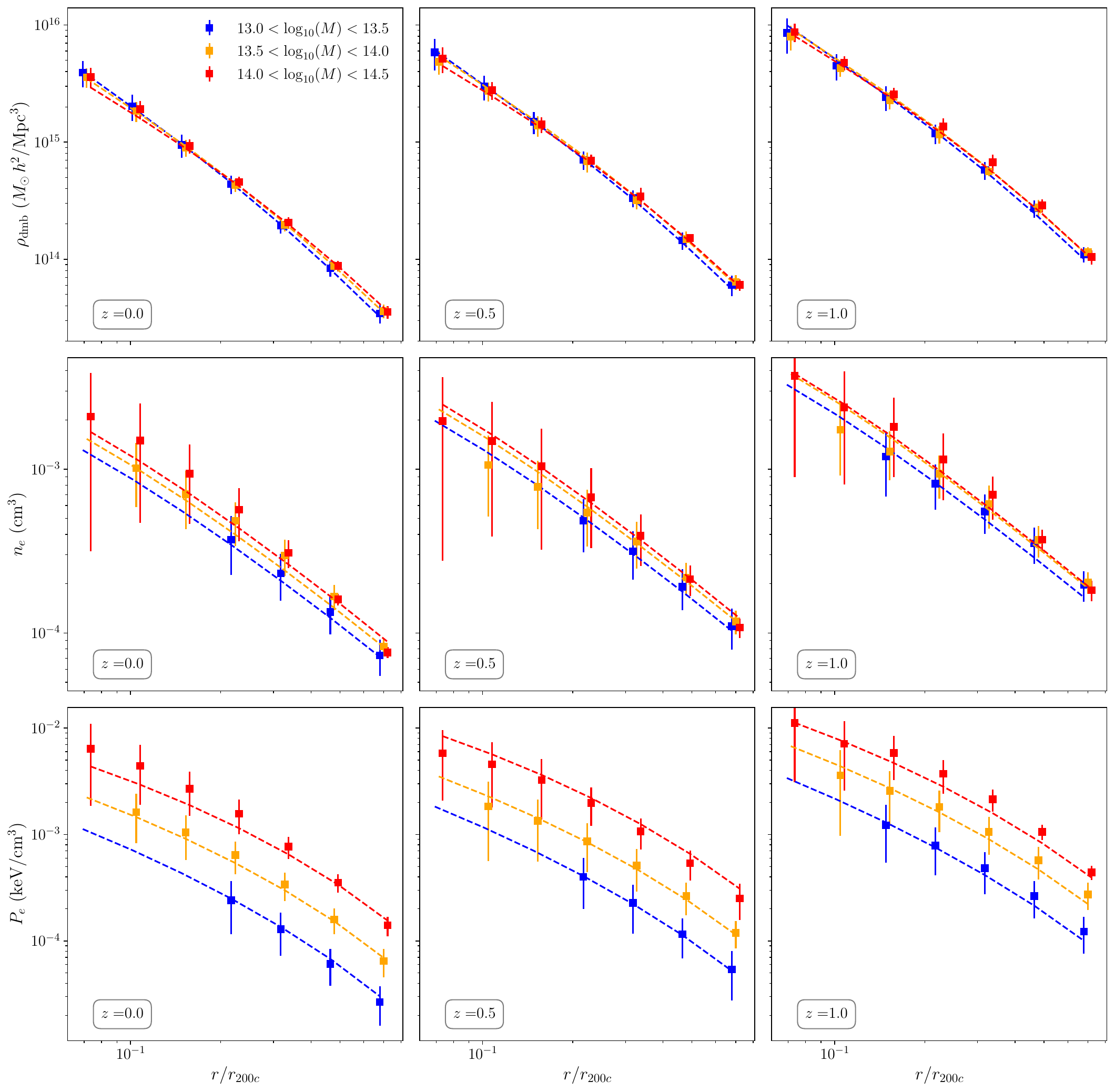}
    \caption{Measurement (markers) and best-fit (dashed lines) of total matter density (top row), electron density (middle row), and electron pressure (bottom row) in one of the ANTILLES simulations. Left, middle, and right columns correspond to snapshots for redshifts 0, 0.5, and 1.0, respectively. Each panel shows the measurement and best-fit for halos in three different mass ranges as mentioned in the legend, going from low to high masses. The markers show the mean and errorbars show the standard deviation of measurement of profiles of all halos in a mass bin. All 27 measurements are fit jointly with the \texttt{GODMAX} code using 10 free parameters (see Section~\ref{sec:fitting_antilles} for details).}
    \label{fig:antilles_fit_example}
\end{figure*}

\section{Results}\label{sec:results}

\begin{figure*}
    \centering
    \includegraphics[width=0.6\linewidth]{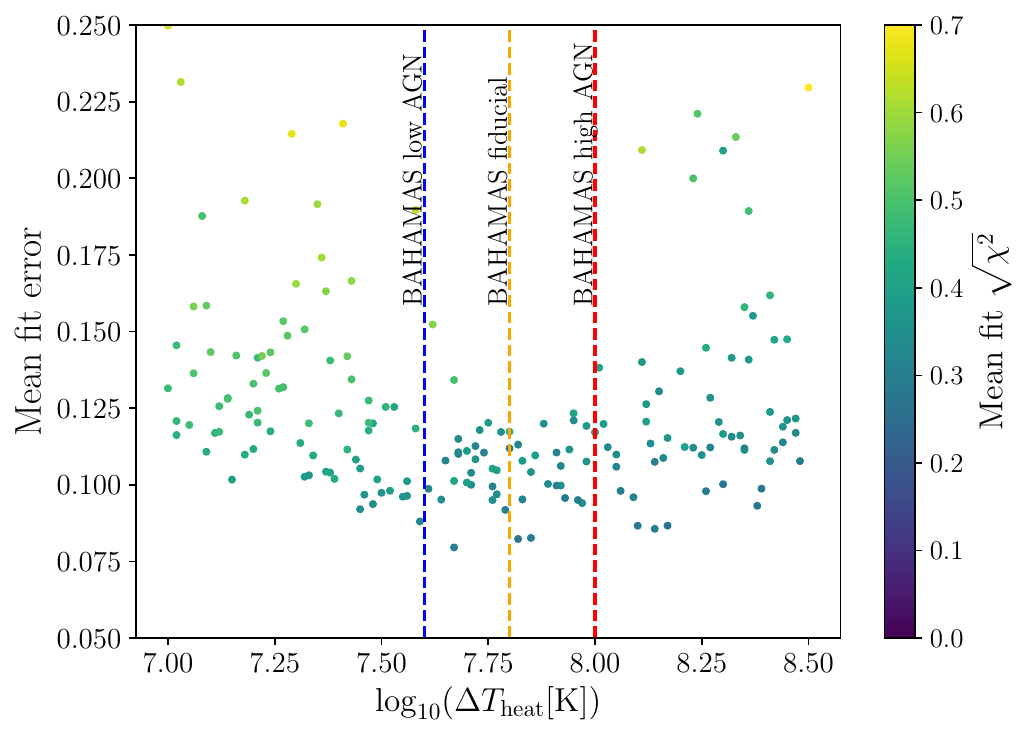}
    \caption{Scatter plot between the goodness of fit for each ANTILLES simulation and their $\Delta T_{\rm heat}$ parameter indicating the energy output of AGN activity. We show the mean errors of the best-fit for all 200 simulations when jointly fitting all 27 profiles shown in Fig.~\ref{fig:antilles_fit_example}. We color each point by the absolute difference between measurement and best-fit relative to their errorbars to show that all the best-fits are consistent with the errorbars. Formally, the mean error of the best-fit is defined as $\langle | (\vec{\xi}_{\rm sim} - \vec{\xi}_{\rm best-fit})/\vec{\xi}_{\rm best-fit}| \rangle$ and mean $\sqrt{\chi^2}$ of the best-fit is defined as $\langle | (\vec{\xi}_{\rm sim} - \vec{\xi}_{\rm best-fit})/\vec{\sigma}_{\rm sim}| \rangle$, where $\vec{\xi}_{\rm sim}$ and $\vec{\xi}_{\rm best-fit}$ are the measurements and best-fit of 27 profiles (see  Fig.~\ref{fig:antilles_fit_example}) respectively and $\vec{\sigma}_{\rm sim}$ is the measurement error. The average is taken over all the measurement points. We also show the values of $\Delta T_{\rm heat}$ used in three \texttt{BAHAMAS} simulation runs, encompassing its expected range while being consistent with the observations. In this expected range of $\Delta T_{\rm heat}$, \texttt{GODMAX} is able to fit the measurements at approximately the 10\% level and in general performs better than the 15\% level. }
    \label{fig:antilles_all_gof}
\end{figure*}

\subsection{Fitting the hydrodynamical simulations}\label{sec:fitting_antilles}

To validate our model, we first focus on reproducing profiles observed in the \texttt{ANTILLES} hydrodynamical simulation suite, specifically targeting three relevant quantities: the {comoving} total matter density ($\rho_{\rm dmb}$), {physical} electron density ($n_e$), and {physical} electron pressure ($P_e$) for halos across various mass and redshift bins. In this validation, we give inputs such as mass, concentration, and redshift of a halo from an N-body simulation and predict $\rho_{\rm dmb}$, $n_e$, and $P_e$ as expected from its paired hydrodynamical simulation. 

We start by dividing halo samples in the \texttt{ANTILLES} N-body simulation snapshots at $z=0.0, 0.5, 1.0$ into three mass bins: $13.0 < \log_{10}(M_{\rm 200c}) < 13.5$, $13.5 < \log_{10}(M_{\rm 200c}) < 14.0$, and $14.0 < \log_{10}(M_{\rm 200c}) < 14.5$. For each snapshot, we perform halo matching between the dark matter-only (DMO) simulation and each of the 200 hydro simulations by finding the closest match in 3D comoving space and halo mass. For each matched halo in the hydro simulation, we calculate $\rho_{\rm dmb}$, $n_e$, and $P_e$ profiles in 7 equally log-spaced radial bins ranging from $0.06 < r/r_{\rm 200c} < 1.0$. Note that we only fit the electron density and pressure originating from non-star forming gas in the halos. We then compute the mean and standard deviation of these profiles for halos within the same mass bin and redshift in the DMO simulation.

Using the 10 free parameters listed in the `baryonic parameters' section of Table~\ref{tab:params_all}, we predict $\rho_{\rm dmb}$, $n_e$, and $P_e$ using the \texttt{GODMAX} code for the specified mass range and redshift values. The analytical predictions for the mean profile are calculated using the halo mass function from the DMO simulation in each mass bin and compared against measurements from each hydro simulation. We employ a bounded minimization of the standard $L_2$ loss using the limited-memory Broyden–Fletcher–Goldfarb–Shanno algorithm \citep{lbfgsb}, utilizing the \texttt{autograd} functionality of \texttt{JAX} to find the best-fit parameters within the ranges mentioned in Table~\ref{tab:params_all}.

In Fig.~\ref{fig:antilles_fit_example}, we present the measurements and best-fit results for one of the \texttt{ANTILLES} simulations. It is evident that we can jointly fit all 27 profiles within the measurement error bars, demonstrating the effectiveness of our model. In Appendix~\ref{app:individual_profiles}, we show the 3D profiles of various individual components corresponding to the bestfit parameters obtained here. 
To quantify the goodness of fit, we employ two statistical measures: mean error and mean $\chi^2$. These results are shown in Fig.~\ref{fig:antilles_all_gof}, where we create a scatter plot that contrasts the average residual between the measurements and the best-fit results against the $\Delta T_{\rm heat}$ parameter. This parameter primarily controls the AGN energy output in each simulation. We find that our bestfits are completely consistent with the measurements under the current uncertainties of the latter. This allows us to verify that the model is accurate to at least 15\% in terms of mean error. Improved measurements of the profiles with larger volume simulations will be needed to determine the model accuracy down to higher precision.
% From our analysis, it is observed that we can fit approximately all simulations to better than 15\% accuracy. 
In Fig.~\ref{fig:antilles_all_gof}, we also include the values of the $\Delta T_{\rm heat}$ parameter for the three large-scale BAHAMAS runs (indicated by vertical lines), which approximately encapsulate the expected range of this parameter, resulting in simulations that are broadly consistent with various observations \citep{McCarthy:2017:MNRAS:}. Within this range, our fits achieve an accuracy of approximately 10\%.

As pointed above, it is important to note that the intrinsic variance in the measurements is large, and our fits are, on average, fully consistent with the measurements. To demonstrate this, we calculate the average absolute difference between the measurements and the best-fit results relative to the measurement error bar. Each point in Fig.~\ref{fig:antilles_all_gof} is colored based on this value, indicating that, on average, our best-fit results are within 50\% of the measurement error. We observe that simulations resulting in comparatively poor fits generally feature either low or high values of the $\Delta T_{\rm heat}$ parameter. With lower $\Delta T_{\rm heat}$ values, fitting the evolution of the pressure profile with redshift becomes challenging. Conversely, at higher $\Delta T_{\rm heat}$ values, the electron density and electron pressure profiles at high redshifts and near the centers of halos tend to become much flatter, leading to poorer fits. It is important to note that these simulations vary across a five-parameter space, each controlling different aspects of baryonic feedback. Consequently, we anticipate a complex interplay between supernovae and AGN efficiency influencing the baryonic profiles \citep{Shao:2023:arXiv:, Gebhardt:2023:arXiv:}. This analysis underscores the robustness of our model in accurately fitting the hydrodynamical simulations across a range of parameters.

In this work, we have concentrated on validating the \texttt{GODMAX} code by directly fitting the radial profiles of various matter component densities and their thermodynamics in halos. The chosen mass range and redshifts of these halos are wide enough to predict the observables such as two-point shear correlation \citep{To:24}, shear-tSZ correlation \citep{Osato:2020:MNRAS:, Pandey:2022}, and kSZ cross-correlations \citep{Amodeo:2021:PhRvD:,Bolliet:2023:JCAP}. However, it is important to note that the tSZ auto-power spectrum is sensitive to significantly higher mass halos at low redshift \citep{Komatsu:2002:MNRAS:,Osato:2020:MNRAS:}, necessitating simulations with substantially larger volumes for accurate reproduction of their statistics \citep{McCarthy:2023:MNRAS:}.\footnote{However, the deeper potential wells of these massive halos tend to reduce the impact of AGN feedback on their pressure profiles, with integrated pressure often aligning with expectations from self-similar scaling relations \citep{Lim2021GasMass, Pop:2022:arXiv:, Lee2022rSZ, Pakmor:2023:MNRAS:}.} We plan to extend the joint validation of the code at the observable level to future studies, utilizing simulations with larger volumes.

\begin{figure*}
    \centering
    \includegraphics[width=1.0\linewidth]{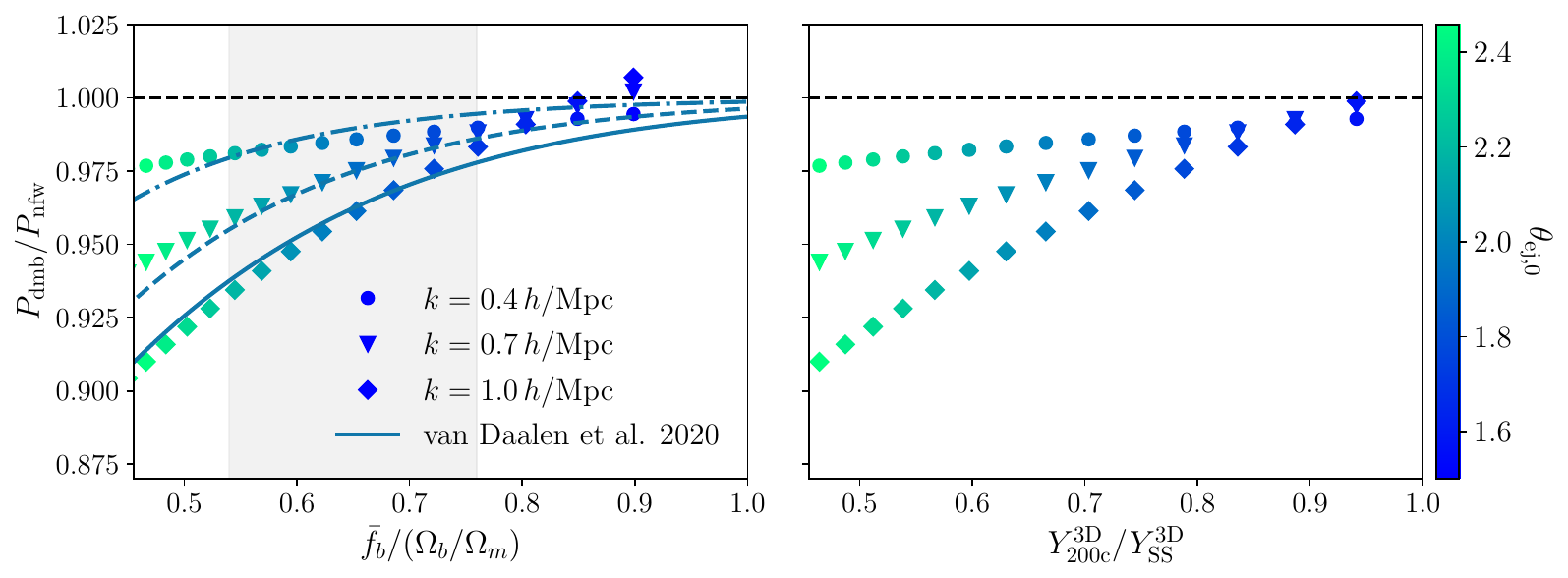}
    \caption{Matter power suppression as a function of mean baryon fraction (left) and mean integrated thermal SZ signal (right) of halos in the mass range $6\times 10^{13} < M_{\rm 200c} [M_{\odot}] < 2\times 10^{14}$. We normalize the baryon fraction with the cosmic baryon fraction and thermal SZ signal with the self-similar scaling. Markers are the predictions obtained from \texttt{GODMAX} code with varying $\theta_{\rm ej,0}$, where higher value of $\theta_{\rm ej,0}$ results in lower values of both $f_{b}$ and $Y^{\rm 3D}_{\rm 200c}$. The lines in the left panel show the fitting function described in \protect\cite{vanDaalen:2020}. Note that the markers are not fit to the lines, rather they reproduce the behavior when changing value of any parameter mimicking the impact of AGN feedback (here $\theta_{\rm ej,0}$). The shaded region in the left panel shows the approximate baryon fraction values consistent with data. In consistency with past studies, we find that matter power suppression is correlated with baryon fraction \protect\citep{vanDaalen:2020} as well as tSZ effect \protect\citep{Pandey:2023:MNRAS:}.}
    \label{fig:Pksup_scaling}
\end{figure*}

\subsection{Constraining the matter power spectrum suppression}

% \begin{itemize}
Recent studies, including \cite{vanDaalen:2020}, have demonstrated that for scales up to $k \sim 1 \, h/{\rm Mpc}$, the normalized baryon fraction ($f_{b}$) of halos with a mass around $10^{14} M_{\odot}$ is a reliable proxy for the total matter power suppression. This finding has been corroborated by several independent suites of simulations \citep{Delgado:2023:MNRAS:, Salcido:2023:MNRAS:, Schaye:2023:MNRAS:}. Furthermore, \cite{Pandey:2023:MNRAS:} has shown that in addition to baryon fraction, the deviation of the normalized integrated tSZ value in high-mass halos from their self-similar expectation ($Y^{\rm 3D}_{\rm 200c}/Y^{\rm 3D}_{\rm SS}$) is also indicative of matter power suppression. These studies used different hydrodynamical simulations with varying feedback strengths and found a consistent correlation: increased AGN feedback tends to expel gas from halos, thereby reducing both $f_{b}$ and $Y^{\rm 3D}_{\rm 200c}$, and consequently diminishing the total matter field power.
% \end{itemize}

As discussed in Section~\ref{sec:Pmm} and Section~\ref{sec:gas_thermodynamics}, our model is capable of calculating matter power suppression, as well as the integrated baryon fraction and tSZ effect, given specific values of baryonic parameters. The parameter $\theta_{\rm ej, 0}$, which effectively controls the ejection radius of gas, serves as a proxy for AGN feedback. Therefore, we run the \texttt{GODMAX} code with varying values of $\theta_{\rm ej, 0}$ within the range $[1.5, 3.0]$ to make predictions for matter power suppression. Additionally, we calculate the mean $f_{b}$ and $Y^{\rm 3D}_{\rm 200c}$ for halos in the same mass range as specified by \cite{vanDaalen:2020} ($6\times 10^{13} < M/M_{\odot} < 2\times 10^{14}$).

In Fig.~\ref{fig:Pksup_scaling}, on the left panel, we plot the power suppression at three different scales ($k \in \{0.4, 0.7, 1.0\} \, h/{\rm Mpc}$) as a function of the mean baryon fraction relative to the cosmic baryon fraction. We overlay the fitting function from \cite{vanDaalen:2020} for these three scales and observe that the \texttt{GODMAX} code produces results consistent with the established relation between power suppression and $f_b$. Notably, the markers are not fitted to the curves from \cite{vanDaalen:2020}; instead, this relation emerges naturally from the analytical framework when varying any parameter that controls the impact of AGN feedback. As described in \cite{vanDaalen:2020}, we also show the approximate range of mean baryon fraction that is consistent with observations for comparison with a gray band \citep{Vikhlinin2006, Maughan2008, Sun2009, Pratt2009, Rasmussen2009, Lin2012, Sanderson2013, Gonzalez2013, Budzynski2014, Lovisari2015, Kravtsov2018, Pearson2017}. Note that, as detailed in \cite{Salcido:2023:MNRAS:}, we expect a similarly tight correlation between $f_b$ and $P_{\rm dmb}/P_{\rm nfw}$ even on smaller scales. We defer a detailed comparison of predictions from the \texttt{GODMAX} code to their fitting to a future study.

While measuring the baryon fraction of halos is straightforward in simulations, doing so in observations it often necessitates processing noisy kSZ or X-ray measurements in the relevant halo mass range of $M \sim 10^{14} M_{\odot}/h$. It is worth noting, however, that this situation is expected to improve with the kSZ cross-correlations enabled by Dark Energy Spectroscopic Instrument (DESI) galaxies \citep{Giri:2022:JCAP:} and X-ray observations by eROSITA \citep{Predehl:2021:A&A:}. Nonetheless, tSZ maps, which are easier to process, serve as sensitive probes of baryon thermodynamics. Current-generation surveys such as the Atacama Cosmology Telescope and South Pole Telescope, with resolutions of approximately 1-2 arcmin, already produce high-resolution tSZ maps~\citep{Madhavacheril:2020:PhRvD:,Coulton:2023,Bleem:2022:ApJS} capable of probing gas pressure inside halos of $M \sim 10^{14} M_{\odot}/h$ at $z < 1$ \citep{Anbajagane:2024:MNRAS:}. Additionally, with upcoming data releases from DESI and SO, we anticipate tight constraints on the pressure profiles of halos around $10^{14} M_{\odot}$ \citep{Pandey:2020}. The right panel of Fig.~\ref{fig:Pksup_scaling} reveals a monotonic correlation between power suppression and $Y^{\rm 3D}_{\rm 200c}$. Consequently, stringent constraints on matter power suppression can be expected using tSZ cross-correlation observations using methodologies as described in \cite{Pandey:2023:MNRAS:, To:24}.

One pertinent question is whether the baryonic model described here can accurately capture the correlation between local SZ measurements and global matter power suppression, thereby maintaining systematic biases in the inferred matter power spectrum under control. \cite{To:24} utilized the mean cluster mass-tSZ scaling of halos to constrain the total matter power spectrum in various hydro simulations. Their findings suggest that SZ information from halos with masses $M_{\rm 200c} \gtrapprox 10^{14} M_{\odot}/h$, as expected from future CMB surveys, is sufficient to ensure that the inferred matter power spectrum is accurate enough. This accuracy is crucial to keep systematic biases in cosmological parameters significantly below the statistical precision of LSST Y1 cosmic shear observations, although they observed larger biases relative to the statistical precision of LSST Y6 observations when using these high-mass cluster samples. We anticipate that analysis of cluster-tSZ correlations with DESI$\times$SO and weak lensing-tSZ correlations with LSST$\times$SO will yield constraints on the pressure profile of lower mass halos, thus reducing systematic uncertainties. A detailed exploration of this effect is reserved for a future study.

% \begin{figure*}
%     \centering
%     \begin{minipage}{0.45\textwidth}
%         \centering
%         \includegraphics[width=\textwidth]{figs/Pksup_fb_thetaej_FINALSET.pdf} % first figure
%         % \caption{First figure}
%     \end{minipage}\hfill
%     \begin{minipage}{0.45\textwidth}
%         \centering
%         \includegraphics[width=\textwidth]{figs/Pksup_Y_th_FINALSET.pdf} % second figure
%     \end{minipage}
%     \caption{Matter power suppression as a function of mean baryon fraction (left) and mean integrated thermal SZ signal (right) of halos in the mass range $6\times 10^{13} < M [M_{\odot}] < 2\times 10^{14}$. We normalize the baryon fraction with the cosmic baryon fraction and thermal SZ signal with the self-similar scaling. Markers are the predictions obtained from \texttt{GODMAX} code with varying $\theta_{\rm ej,0}$, where higher value of $\theta_{\rm ej,0}$ results in lower values of both $f_b$ and $Y_{\rm 200c}$. Solid lines in the left panel show the fitting function described in \cite{vanDaalen:2020}. Note that the markers are not fit to the solid lines, rather they reproduce the behavior when changing value of any parameter mimicking the impact of AGN feedback (here $\theta_{\rm ej,0}$). In consistency with past studies, we find that matter power suppression is correlated with baryon fraction \citep{vanDaalen:2020} as well as tSZ effect \citep{Pandey:2023:MNRAS:}. }    
%     \label{fig:Pksup_scaling}
% \end{figure*}

% \subsection{Dependence of SZ on secondary halo properties}

\begin{figure*}
    \centering
    \includegraphics[width=1.0\linewidth]{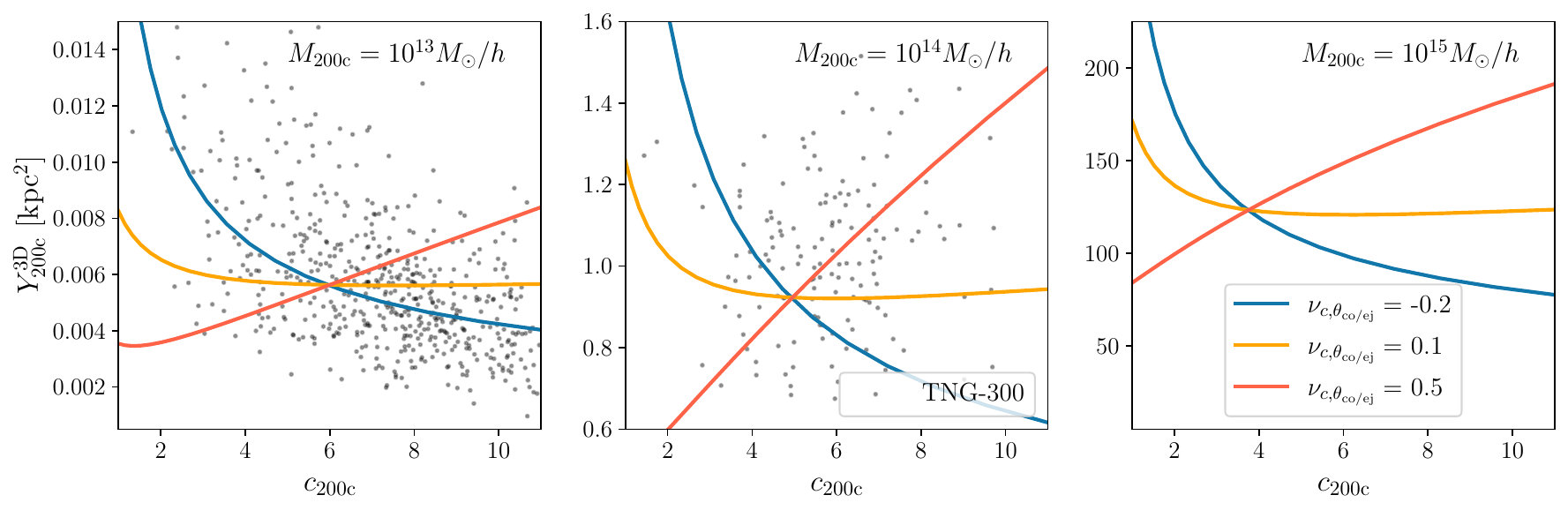}
    \caption{Correlation between integrated tSZ signal and halo concentration and its dependence on the gas density concentration for three different halo masses as given in plot inset. Each panel shows this correlation for three different values of $\eta_{\rm ej/co}$, which controls the evolution of $\theta_{\rm co}$ and $\theta_{\rm ej}$ parameters with halo concentration (see Eq.~\ref{eq:theta_ej_co}), with higher values signifying higher correlation between gas density concentration and dark matter density concentration of the halo. We also show the rescaled measurement of $Y^{\rm 3D}_{\rm 200c} - c_{\rm 200c}$ relation in the Illustris-TNG simulation in the first two panels (see Section~\ref{sec:Y-c_relation} for details).}
    \label{fig:Y_c_relation}
\end{figure*}

\subsection{Dependence of SZ on secondary halo properties}\label{sec:Y-c_relation}

The formation of a halo is a complicated process where its evolution depends upon its formation history and environment. Consequently, the hydrodynamical properties of gas within a halo can depend on secondary halo properties beyond mass. Halo concentration is one such property, acting as a tracer of halo formation and accretion history. Studies such as \cite{Lee:2022:MNRAS:}, \cite{Wadekar:2022}, \cite{Baxter:2024:MNRAS:}, and \cite{Hadzhiyska:2023:MNRAS:} have shown that both the pressure and density of hot gas are sensitive to halo concentration. These analyses, employing different simulations and pressure profile models, find that for high-mass clusters ($M > 10^{14} M_{\odot}$), the 3D integrated tSZ signal ($Y^{\rm 3D}_{\rm 200c}$, see Eq.~\ref{eq:Y3D}) positively correlates with halo concentration. This correlation is expected in high-mass clusters with deep potential wells, where the concentration of hot gas tends to mirror that of dark matter. Assuming hydrostatic equilibrium, the pressure profile within the halo radius would also be more concentrated. However, in lower-mass halos, AGN activity can disrupt this correlation due to their shallower potential wells, leading to more gas in the outskirts of the halo \citep{Hadzhiyska:2023:MNRAS:}.

In Fig.~\ref{fig:Y_c_relation}, we illustrate the relationship between $Y^{\rm 3D}_{\rm 200c}$ (see Eq.~\ref{eq:Y3D}) and $c_{\rm 200c}$ for halos across three different halo masses: $M_{\rm 200c} \in \{10^{13}, 10^{14}, 10^{15} \} M_{\odot}/h$. We observe this relationship for varying values of $\eta_{\rm ej}$ and $\eta_{\rm co}$, where a higher value indicates a stronger correlation between the concentration of the gas profile and the dark matter profile. We find that for higher values of $\eta_{\rm ej/co}$, there is a positive correlation between $Y^{\rm 3D}_{\rm 200c}$ and $c_{\rm 200c}$, whereas for lower values, the correlation weakens or even reverses.

We compare the predictions of the $Y^{\rm 3D}_{\rm 200c} - c_{\rm 200c}$ relationship to measurements in the 300 Mpc box of the Illustris-TNG simulation \citep{Nelson:2019:ComAC:}. 
% Several simplifying assumptions are made in these measurements, as our goal in this subsection is to illustrate the qualitative trends of changes in the $Y^{\rm 3D}_{\rm 200c} - c_{\rm 200c}$ relationship with halo masses due to variations in baryonic feedback, as captured by the \texttt{GODMAX} code. Firstly, $c_{\rm 200c}$ is measured using only the dark matter particles from the hydro simulations, rather than the matched halo catalog from the gravity-only simulation, as required by the analytic code. However, this is deemed a sufficient approximation since the dark matter profile in hydro simulations remains close to the NFW profile, even after the adiabatic relaxation discussed in Section~\ref{sec:clm_profile}. Next, to reduce memory requirements when measuring the pressure profile and $Y^{\rm 3D}_{\rm 200c}$ of halos, only gas particles identified as being inside the halo using the Friends-of-Friends algorithm are used. This approach may miss capturing some gas particles near the outskirts of the halo, potentially leading to an underestimation of the measured $Y^{\rm 3D}_{\rm 200c}$. However, we do not expect this suppression to be dependent on halo concentration. Therefore, the measured $Y^{\rm 3D}_{\rm 200c}$ values are scaled to approximately align with the analytic predictions shown in Fig.~\ref{fig:Y_c_relation} at the mean concentration for each halo mass. 
We observe that for halos with a mass of $10^{13} \, M_{\odot}/h$, $Y^{\rm 3D}_{\rm 200c}$ and $c_{\rm 200c}$ are negatively correlated, which is expected as the higher concentration halos result in more efficient AGN formation in the central galaxy. This leads to an increased baryonic feedback which pushes the gas to the halo outskirts and hence reducing its thermal pressure. In contrast, halos with higher masses have deeper potential wells which reduces the impact of baryonic feedback \citep{Hadzhiyska:2023:MNRAS:}. This results in a positive correlation between gas and dark matter concentration which leads to higher thermal pressure in halos with higher concentration. We observe a moderately positive correlation between $Y^{\rm 3D}_{\rm 200c}$ and $c_{\rm 200c}$ in halos with mass of approximately $10^{14} \, M_{\odot}/h$. This analysis suggests the need for a mass-dependent evolution of the $\eta_{\rm ej/co}$ parameter to capture the evolution of $Y^{\rm 3D}_{\rm 200c} - c_{\rm 200c}$ relationship with halo mass. Due to the limited halo samples in the \texttt{ANTILLES} suite within the relevant mass range of $10^{13} < M [M_{\odot}/h] < 10^{14.5}$ for creating a split in both halo mass and concentration, a careful calibration of this effect is deferred to future studies using simulations with larger volumes.

\begin{figure*}
    \centering
    \includegraphics[width=1.0\linewidth]{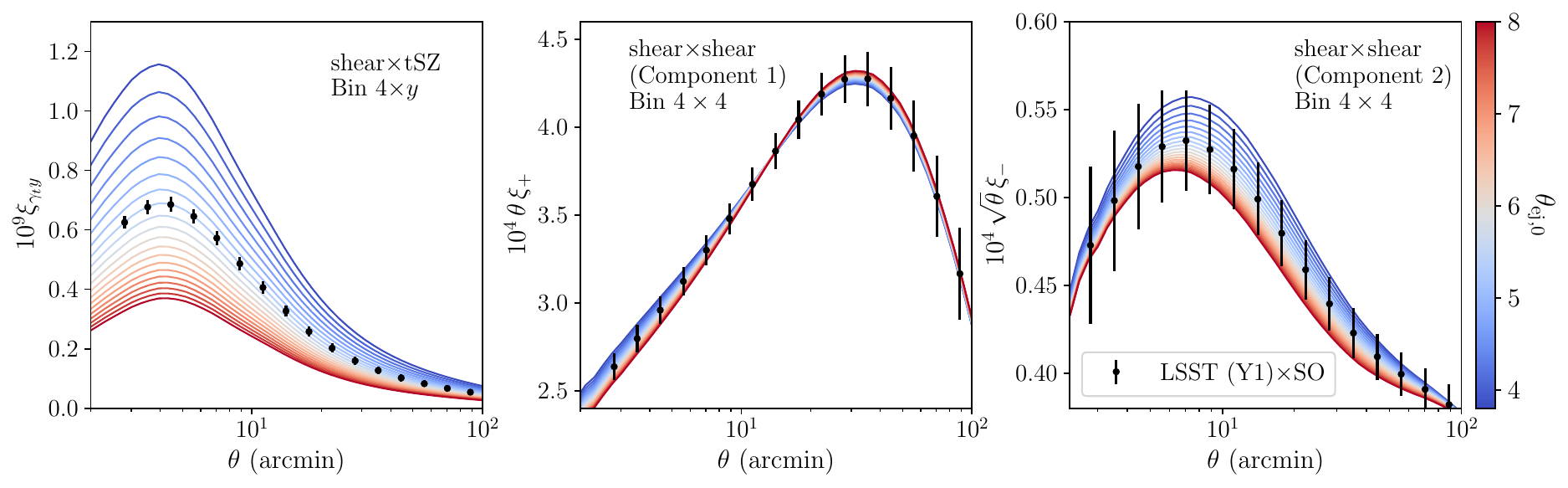}
    \caption{Simulated measurement of correlations between weak lensing and thermal SZ effect from LSST Y1 and SO respectively. Left panel shows cross-correlation between weak lensing and thermal SZ effect whereas middle and right panel correspond to weak lensing auto-correlations, $\xi_{+}$ and $\xi_{-}$. We show the simulated measurement and errorbars for the fourth tomographic source bin of LSST. We also show the simulated curves with varying value of $\theta_{\rm ej,0}$ parameter.}
    \label{fig:forecast_dv_example}
\end{figure*}

\begin{figure*}
    \centering
    \includegraphics[width=1.0\linewidth]{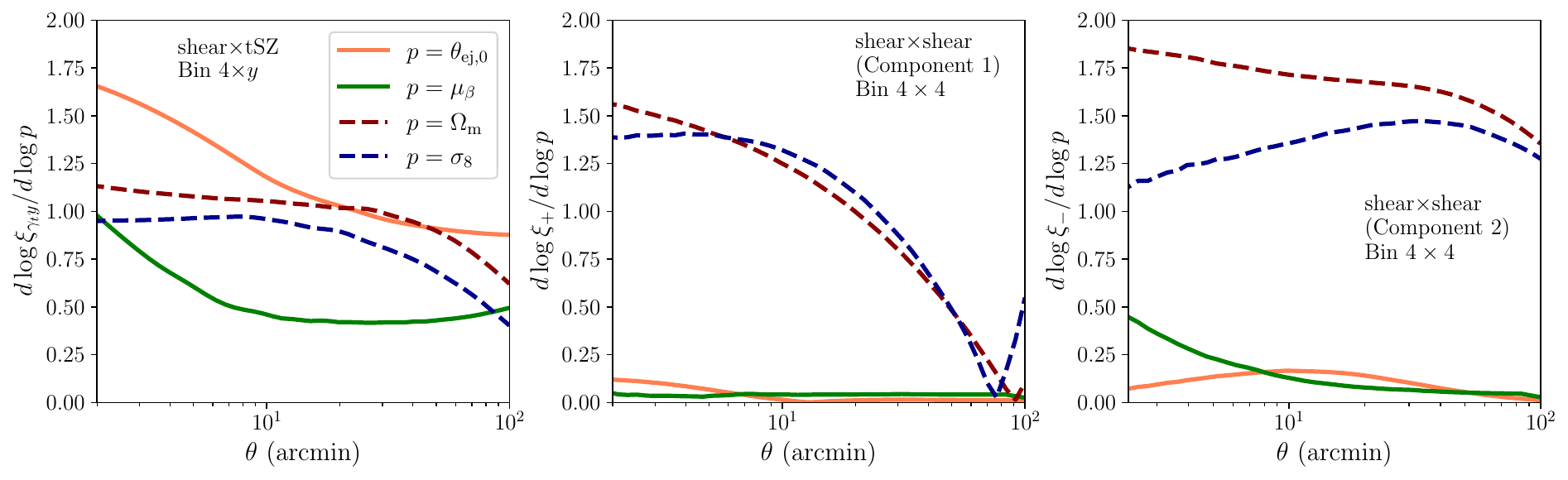}
    \caption{Sensitivity of $\xi_{\gamma_t y}$ (left), $\xi_{+}$ (center), and $\xi_{-}$ (right) to baryonic (solid lines) and cosmological parameters (dashed lines). We can see that $\xi_{\gamma_t y}$ is significantly more sensitive to baryonic parameters compared to shear alone. Note that the plots are for correlations with the fourth source tomographic bin. }
    \label{fig:sensitivity_of_dv}
\end{figure*}

\begin{figure*}
    \centering
    \begin{minipage}{0.45\textwidth}
        \centering
        \includegraphics[width=\textwidth]{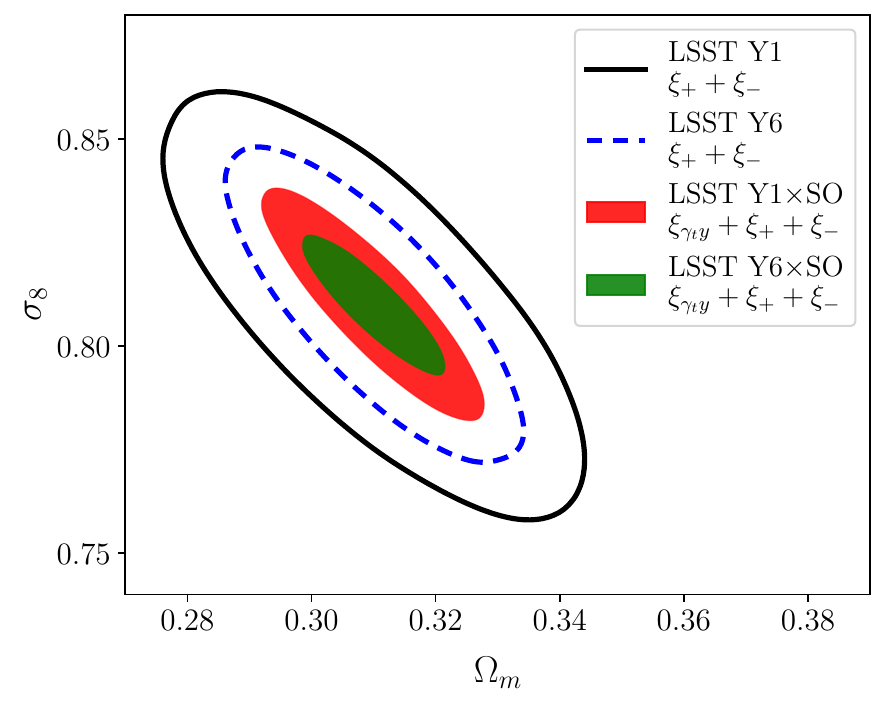} % first figure
        % \caption{First figure}
    \end{minipage}\hfill
    \begin{minipage}{0.45\textwidth}
        \centering
        \includegraphics[width=\textwidth]{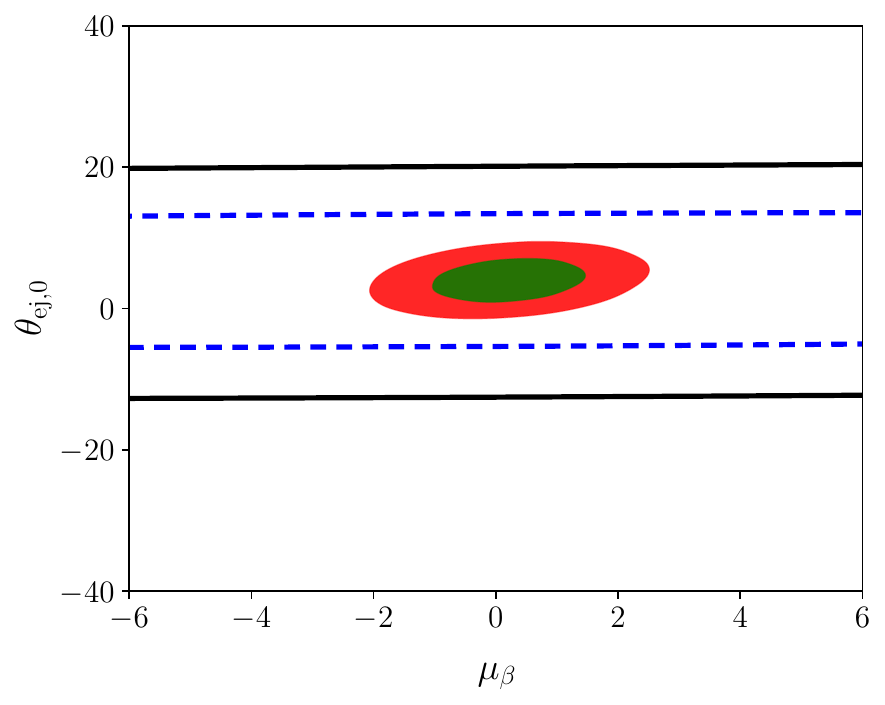} % second figure
    \end{minipage}
    \caption{Forecasted constraints on cosmological (left) and gas distribution (right) parameters when analyzing different cosmic probe combinations as described in the legend. The unfilled contours show the constraints with only the weak lensing auto-correlations, while the filled contours show the results for a joint analysis that also includes weak lensing-tSZ cross-correlations. We see that including the cross-correlations significantly improves the constraints on the gas distribution parameters and also improves cosmological parameter constraints by breaking their degeneracies with the baryonic parameters.}   
    \label{fig:forecast_constraints}
\end{figure*}

\subsection{Forecast for future LSS$\times$CMB surveys}\label{sec:forecast}
We now turn our attention to forecasting the baryonic and cosmological constraints expected from the upcoming LSST weak lensing and SO tSZ observations. Our analysis jointly examines the two-point correlations $\xi_{\gamma_t y}$, $\xi_{+}$, and $\xi_{-}$. We calculate the joint covariance (as detailed in Section~\ref{sec:covariance}) incorporating the noise expected from both LSST Y1 and LSST Y6 source catalog number densities, as well as SO tSZ maps (refer to Section~\ref{sec:forecast_specification} for noise specifications). Notably, the simulated source catalog is divided into five tomographic bins (illustrated in Fig.\ref{fig:nz_sources}).

In Fig.~\ref{fig:forecast_dv_example}, we present the simulated data vector for the fourth source tomographic bin, along with error bars corresponding to the LSST Y1$\times$SO analysis. We also demonstrate the variation resulting from altering a single parameter, $\theta_{\rm ej, 0}$, which controls the truncation radius of the gas profile relative to $r_{\rm 200c}$. The parameter range illustrated in the plot aligns with the best-fit $\theta_{\rm ej, 0}$ values obtained from the \texttt{ANTILLES} simulation suite. We observe that variations in $\theta_{\rm ej, 0}$ lead to changes in the weak lensing auto-correlations $\xi_{+}$ and $\xi_{-}$ at approximately $1\sigma$ of the measured error bars, especially at smaller scales. This variability limits our ability to extract cosmological information from these scales. However, the cross-correlation function $\xi_{\gamma_t y}$ exhibits a significantly higher sensitivity to changes in $\theta_{\rm ej, 0}$.

To explicitly assess the sensitivity of the two-point correlations to various parameters, we can examine their log-derivatives. Leveraging the auto-differentiation property of the \texttt{GODMAX} model within the \texttt{JAX} framework, we can calculate exact derivatives of the final data vector with respect to any input parameter. In Fig.~\ref{fig:sensitivity_of_dv}, we display the log-derivatives of $\xi_{\gamma_t y}$, $\xi_{+}$, and $\xi_{-}$ with respect to two cosmological parameters ($\Omega_{\rm m}$ and $\sigma_8$), as well as two baryonic parameters ($\theta_{\rm ej, 0}$ and $\mu_{\beta}$). All three correlations exhibit high sensitivity to the cosmological parameters. However, as anticipated, $\xi_{\gamma_t y}$ shows a notably higher sensitivity to baryonic parameters. Consequently, we expect that a joint analysis of all three correlations will facilitate the constraining of both baryonic and cosmological parameters effectively.

In Fig.~\ref{fig:forecast_constraints}, we present the constraints on both cosmological and baryonic parameters derived from the Fisher forecast analysis (as detailed in Section~\ref{sec:fisher_forecast}). This forecast utilizes the complete data vector of $\xi_{\gamma_t y}$, $\xi_{+}$, and $\xi_{-}$, constructed from all five source tomographic bins and the tSZ map (outlined in Section~\ref{sec:datavector_modeling}) between the angular scales of 2.5 and 250 arcmin. Additionally, we vary all parameters listed in Table~\ref{tab:params_all}, applying Gaussian priors on shear calibration and source photometric redshift biases, while adopting wide, uninformative priors for all other parameters.

The unfilled contours in the figure illustrate the expected constraints achievable by analyzing weak lensing auto-correlations alone, considering the noise levels anticipated from LSST Y1 and LSST Y6 source galaxies. The results indicate that, as expected, LSST Y6 observations will yield more precise constraints in both cosmological and baryonic parameters compared to Y1. We calculate the 2D figure of merit (FoM), defined for any two parameters $p_1$ and $p_2$ as ${\rm FoM}_{p_1, p_2} = \sqrt{{\rm det[\mathcal{C}(p_1,p_2)]}}$, where $\mathcal{C}(p_1,p_2)$ is the covariance of marginalized constraints on the two parameters. We find that going from Y1 to Y6, the FoM of $\Omega_{\rm m}-\sigma_8$ improves from 1906 to 3994, giving a factor of 2.1 improvement in cosmological constraining power. 

However, the filled contours represent the enhanced precision in constraints achievable through a joint analysis of all three data vectors $\xi_{\gamma_t y}$, $\xi_{+}$, and $\xi_{-}$. This combined approach significantly sharpens the precision of constraints, particularly for baryonic parameters, which then breaks degeneracies with the cosmological parameters. In the two cosmological parameters, $\Omega_{\rm m}-\sigma_8$, the FoM improves by a factor of 4.8 and 6.2 for Y1 and Y6 respectively when including $\xi_{\gamma_t y}$ in the datavector. Note that $\xi_{+}$ and $\xi_{-}$ are forecasted to be measured with significantly higher signal-to-noise (approximately 235 and 380 for Y1 and Y6 respectively) compared to $\xi_{\gamma_t y}$ (80 and 136 for Y1 and Y6 respectively) alone. Even with relatively smaller signal-to-noise, the cross-correlation function $\xi_{\gamma_t y}$ plays a crucial role in breaking the degeneracies between cosmological and baryonic parameters, leading to more robust and precise constraints across the full parameter space. This finding underscores the significant benefits of conducting a joint analysis of weak lensing and tSZ data.

\section{Conclusions}\label{sec:summary}
The analysis of Large Scale Structure (LSS) and Cosmic Microwave Background (CMB) observations has entered an era where theoretical predictions' accuracy, rather than measurement uncertainties, often limits the results. To date, most weak lensing analyses have relied on predictions based on gravity-only simulations, with some analyses allowing for marginalization over baryonic effects. However, as current and upcoming analyses aim to probe smaller scales and higher-order statistics, a more nuanced understanding of baryonic effects becomes crucial.

The thermal Sunyaev-Zel'dovich (tSZ) and kinetic Sunyaev-Zel'dovich (kSZ) effects offer direct probes of the baryon distribution and their thermodynamics. Therefore, a joint analysis of weak lensing and SZ effects can significantly aid in concurrently constraining cosmological and baryonic parameters. In this work, we introduce the \texttt{GODMAX} model, designed to describe both weak lensing and SZ effects, along with their correlations. This model facilitates their joint analysis, thus bridging a critical gap in current theoretical frameworks and enhancing our understanding of the Universe's structure and evolution.

A summary of our main results is as follows:
\begin{itemize}
    \item We have developed a physical model that jointly predicts the distribution of dark matter, baryons, and baryonic thermodynamics. This flexible formalism, based on minimal assumptions, is specifically designed for the joint analysis of correlations between weak lensing and both the kinematic and thermal Sunyaev-Zel'dovich effects.

    \item The model has been validated using a suite of 200 \texttt{ANTILLES} hydro simulations featuring a wide range of feedback implementations. It successfully jointly fits the total matter density profiles, electron density profiles, and electron pressure profiles for halos within $0 < z < 1$ and $13 < \log_{10}(M_{\rm 200c}) < 14.5$, achieving an approximate 10\% accuracy level and full consistency within measurement error bars (refer to Fig.~\ref{fig:antilles_all_gof}).

    \item The model accurately reproduces the scaling relations observed in hydro simulations between local halo baryon fraction and total matter power suppression without any fine-tuning (see left panel of Fig.~\ref{fig:Pksup_scaling}). Similarly, it demonstrates how local integrated SZ signals correlate with total matter power suppression (see right panel of Fig.~\ref{fig:Pksup_scaling}).

    \item Our analysis shows that compared to a weak lensing-only analysis, a joint analysis incorporating weak lensing and tSZ cross-correlations yields significantly improved constraints on both cosmological and baryonic distribution parameters (see Fig.~\ref{fig:forecast_constraints}). This improvement is primarily due to the higher sensitivity of weak lensing $\times$ tSZ cross-correlations to baryonic parameters (as shown in Fig.~\ref{fig:sensitivity_of_dv}).
    
    \item The entire codebase is developed using the \texttt{JAX} library, featuring \texttt{autodiff} and \texttt{jit} functionality along with GPU-ready compilation of all correlations and likelihood computations. The entire pipeline to calculate the likelihood for simulated observations of $\xi_{\gamma_t y}$, $\xi_{+}$, and $\xi_{-}$ for the LSST$\times$SO analysis described in Section~\ref{sec:fisher_forecast} takes approximately 15 seconds. The differentiable nature of the likelihood also facilitates integration with Hamiltonian Monte-Carlo sampling schemes, hence enabling efficient sampling of parameter space.
\end{itemize}

The modular nature of the \texttt{GODMAX} code allows for several future enhancements. Our immediate aim is to integrate CMB lensing as an additional probe. CMB lensing, being sensitive to the total matter distribution with a mean redshift sensitivity around $z\sim 2$, will significantly broaden the redshift range accessible in joint analyses. Moreover, while the forecasts presented in this work have concentrated on tSZ and weak lensing cross-correlations, the capability of \texttt{GODMAX} to jointly predict gas density profiles opens up the possibility of incorporating kSZ and X-ray observations into our modeling framework as well. As the calculations are based on halo model, we can additionally incorporate galaxy distributions using the halo occupation distribution framework \citep{Berlind:2003:ApJ:}. This will pave the way for jointly analyzing the correlations of galaxies with gas thermodynamics and matter distribution as well. 

Furthermore, with a comprehensive prescription of SZ profiles, we can employ these models to `baryonify' existing N-body simulations. This process involves correcting the total matter distribution for baryonic effects \citep{Schneider:2016:JCAP:} and pasting consistent gas density and pressure profiles onto them (see \citealt{Anbajagane:24}). Such an approach enables the analysis of higher-order statistics in weak lensing and SZ observables, potentially leading to more refined constraints on feedback mechanisms and cosmological parameters.

\section{Acknowledgements}
We thank Elisabeth Krause, Fran\c{c}ois Lanusse, Daisuke Nagai, Greg Bryan, Learning the Universe and Baryon Pasters groups for useful discussions and comments on the draft. SP is supported by the Simons Collaboration on Learning the Universe.  This project has received funding from the European Research Council (ERC) under the European Union's Horizon 2020 research and innovation programme (grant agreement No 769130). JCH acknowledges support from NSF grant AST-2108536, NASA grants 21-ATP21-0129 and 22-ADAP22-0145, the Sloan Foundation, and the Simons Foundation. EB is partially supported by NSF Grant AST-2306165. This is not an official SO Collaboration paper. 

\section{Data Availability}
The code will be made publicly available at \url{https://github.com/shivampcosmo/GODMAX} upon acceptance of the paper by a journal. The data underlying this article will be shared upon reasonable request to the corresponding authors.

\bibliographystyle{mnras}
\bibliography{thebib,ads}

\appendix

\begin{figure*}
    \centering
    \includegraphics[width=1.0\linewidth]{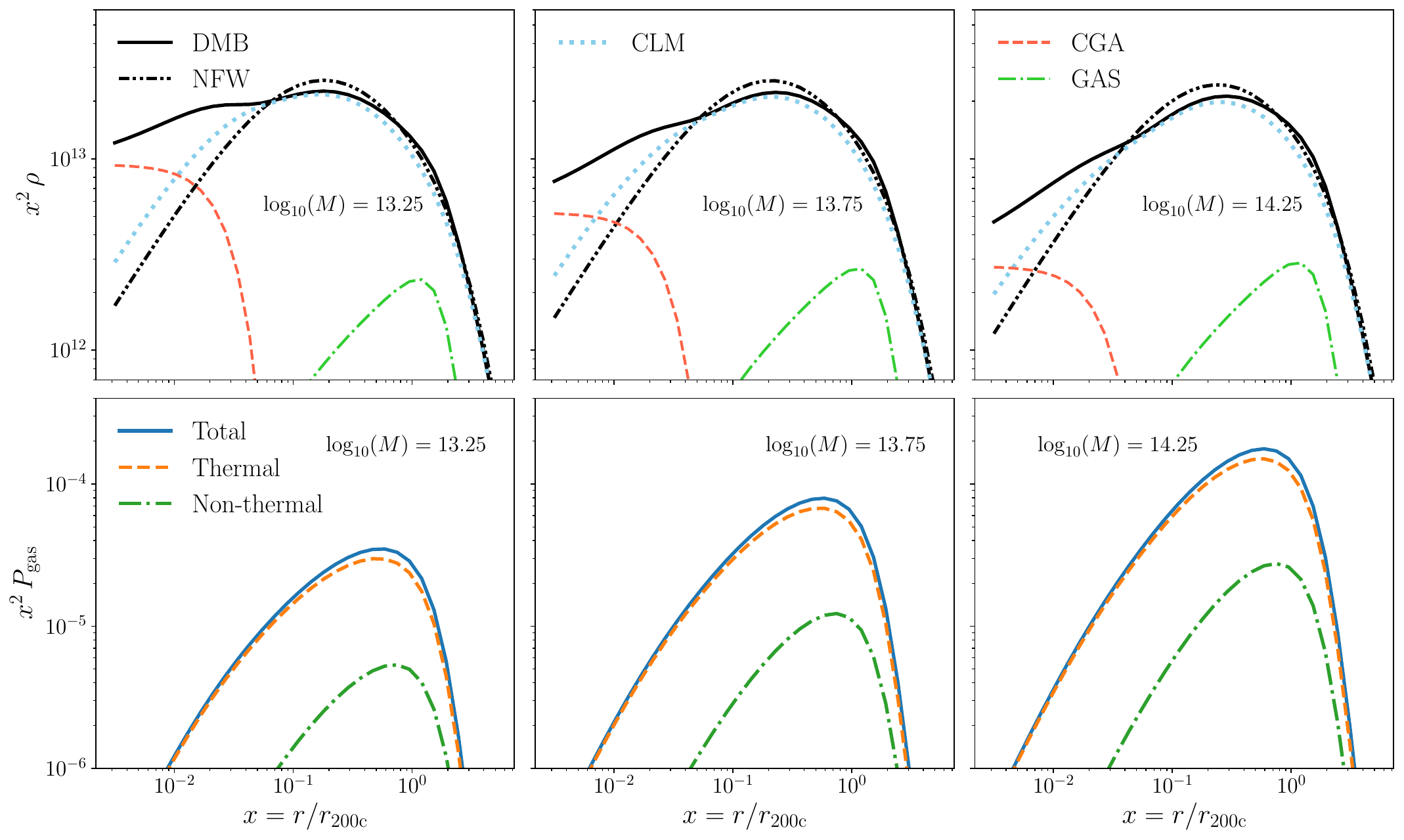}
    \caption{Density profiles of individual matter components in the DMB model, as compared to NFW profile (top row) and pressure of hot gas (bottom row) at $z=0$. We show the profiles for halos of three different mass values in different columns as given in the inset text. In top row, the solid black line corresponds to profile of total matter (DMB) including the contributions from stars in central galaxy (CGA), collisionless matter (CLM) and baryonic gas (GAS). We also show the NFW profile for halo with same mass using black dash-dot-dot-dashed line. The bottom row shows the predictions of total pressure including the contributions from thermal and non-thermal components. The profiles are generated at the parameter values corresponding the the bestfit model shown in Fig.~\ref{fig:antilles_fit_example}.}
    \label{fig:individual_components}
\end{figure*}
\section{Profiles of individual components}\label{app:individual_profiles}
In Fig.~\ref{fig:individual_components}, we display the profiles of individual components within the DMB model (Eq.~\ref{eq:rho_dmb}) of matter components, juxtaposed with the NFW profile (Eq.~\ref{eq:rho_nfw}) expected from halos of same mass at $z=0$. As detailed in Section~\ref{sec:model}, the total matter (DMB) profile can be decomposed into three major components: the central galaxy (CGA), collisionless matter (CLM), and baryonic gas (GAS). These profiles are generated using the parameter values corresponding to the best-fit \texttt{GODMAX} model for the simulation measurements shown in Fig.~\ref{fig:antilles_fit_example}. Our findings indicate that stars in the central galaxy predominantly influence the profile at small scales. At intermediate scales, the total matter profile appears flatter compared to the NFW profile, attributable to the gas being ejected out of the halo towards the outskirts. Additionally, we present a consistent prediction of the pressure of hot gas, including contributions from both thermal and non-thermal components. Notably, we find that non-thermal support comprises a non-negligible fraction of the total pressure, particularly at the outskirts of the halo. Note that the bestfit parameters with which these profiles are generated are $\theta_{\rm ej, 0}=2.08$, $\theta_{\rm co, 0}=0.0032$, $\nu_{\rm M_{\rm c}}=-5.9$, $\mu_{\rm ej}=3.45$, $\gamma=4.143$, $\mu_{\beta}=0.014$, $\log_{10}(M_{\rm c,0})=14.25$, $\mu_{\rm star}=0.19$, $\alpha_{\rm nt}=0.17$, $\nu_{\rm ej}=0.001$.

\bsp    % typesetting comment
\label{lastpage}
\end{document}